\documentclass[fleqn,usenatbib]{mnras}

\usepackage{newtxtext,newtxmath}

\usepackage[T1]{fontenc}
\usepackage{ae,aecompl}
\usepackage{float}
\RequirePackage{fix-cm}

\usepackage[normalem]{ulem}
\usepackage{graphicx}	
\usepackage{amsmath}	
\usepackage{amssymb}	
\usepackage{caption}
\usepackage{subcaption}

\usepackage{hyperref}

\newcommand{\cm}{\,cm\,}
\newcommand{\km}{\,km\,}
\newcommand{\pc}{\,pc\,}
\newcommand{\kpc}{\,kpc\,}
\newcommand{\s}{\,s\,}
\newcommand{\Myr}{\,Myr\,}
\newcommand{\Gyr}{\,Gyr\,}
\newcommand{\kms}{\,km\,s$^{-1}$\,}
\newcommand{\muG}{\,$\mu$G\,}
\newcommand{\nG}{\,$10^{-9}$G\,}
\newcommand{\g}{\,g\,}
\newcommand{\pcc}{\,cm$^{-3}$\,}

\newcommand{\emf}{\mbox{\boldmath ${\cal E}$} {}}
\newcommand{\mean}[1]{\overline{#1}}

\newcommand{\eref}{Eq.~\,\ref}
\newcommand{\fref}{Fig.~\,\ref}
\newcommand{\sref}{Sec.~\,\ref}
\newcommand{\aref}{Appendix~\,\ref}

\usepackage{color,ulem,soul}
\definecolor{dark-red}{rgb}{0.75, 0.00, 0.00}
\definecolor{magenta}{rgb}{0.75, 0.25, 0.75}

\definecolor{hlcolor}{rgb}{1.00, 0.90, 0.85}\sethlcolor{hlcolor}

\title[Dynamo Coefficients in ISM turbulence via SVD]{Turbulent transport coefficients in galactic dynamo simulations using singular value decomposition}

\author[A. Bendre et al.]{
Abhijit B. Bendre,$^{1}$\thanks{E-mail: abhijit@iucaa.in}
Kandaswamy Subramanian,$^{1}$\thanks{E-mail: kandu@iucaa.in}
Detlef Elstner$^{2}$\thanks{E-mail: elstner@aip.de}
and Oliver Gressel$^{2}$\thanks{E-mail: ogressel@aip.de}
\\
$^{1}$IUCAA, Post Bag 4, Ganeshkhind, Pune 411007, India\\
$^{2}$Leibniz-Institut f{\"u}r Astrophysik Potsdam (AIP), An der Sternwarte 16, 14482 Potsdam, Germany
}

\date{Accepted XXX. Received YYY; in original form ZZZ}

\pubyear{2015}

\begin{document}
\label{firstpage}
\pagerange{\pageref{firstpage}--\pageref{lastpage}}
\maketitle

\begin{abstract}
	Coherent magnetic fields in disc galaxies are thought 
	to be generated by a large-scale (or mean-field) dynamo 
	operating in their interstellar medium. A key driver of 
	mean magnetic field growth is the turbulent electromotive 
	force (EMF), which represents the influence of correlated 
	small-scale (or fluctuating) velocity and magnetic fields 
	on the mean field. The EMF is usually expressed as a 
	linear expansion in the mean magnetic field and its 
	derivatives, with the dynamo tensors as expansion 
	coefficients. Here, we adopt the singular value 
	decomposition (SVD) method to directly measure these 
	turbulent transport coefficients in 
	a simulation of the turbulent interstellar medium that 
	realizes a large-scale dynamo. Specifically, the SVD is used 
	to least-square fit the time series data of the EMF with 
	that of the mean field and its derivatives, to determine 
	these coefficients. We demonstrate that the spatial profiles 
	of the EMF reconstructed from the SVD coefficients match well 
	with that taken directly from the simulation. Also, as a
	direct test, we use the coefficients to simulate a 1-D 
	mean-field dynamo model and find an overall similarity in the 
	evolution of the mean magnetic field between the dynamo model 
	and the direct simulation. We also compare the results with 
	those which arise using simple regression and the ones 
	obtained previously using the test-field (TF) method, 
	to find reasonable qualitative agreement. Overall, the 
	SVD method provides an effective post-processing tool to 
	determine turbulent transport coefficients from simulations.
\end{abstract}

\begin{keywords}
galaxies: magnetic fields -- dynamo -- ISM: magnetic fields -- (magnetohydrodynamics) MHD -- methods: data analysis -- turbulence 
\end{keywords}



\section{Introduction}
\label{intro}

Magnetic fields hosted by the nearby spiral galaxies 
are observed to have a coherent large-scale component 
spanning kilo-parsec length scales, and strengths of 
several micro-Gauss 
\citep{fletcher_nearby_2010,Beck2012,beck_wielebinski,krause2018chang}. 
Such large-scale magnetic fields are thought to be 
maintained by a mean-field or large-scale dynamo 
through the combined action of helical interstellar 
turbulence and galactic differential rotation 
\citep[see eg.][and references therein]{beck_1996,shukurov_2005,anvar_2004}. 
The mathematical modelling of the large-scale dynamo, 
relies upon mean-field electrodynamics, where the 
magnetic field $\mathbf{B}$ is split into a mean field 
$\mean{\mathbf{B}}$ and a fluctuation $\mathbf{b}$ and 
similarly for the velocity field $\mathbf{U}=\mean{
\mathbf{U}} +\mathbf{u}$, with the mean defined by some 
suitable averaging \citep{Mof78,BS05}.

The averaged induction equation then picks up a new 
contribution, the mean turbulent EMF $\mean{\mathbf{
\mathcal{E}}}= \mean{ \mathbf{ u } \times \mathbf{b}} 
$ which is the cross correlation between fluctuating 
velocity and magnetic field and is crucial for driving 
the large-scale dynamo. In order to get a closed 
equation for the mean magnetic field, using a two-scale 
approach, $\mean{\mathbf{\mathcal{E}}} $ is expressed 
as a linear expansion in the mean magnetic field and 
its derivatives. The resulting expansion coefficients 
encapsulate the various properties of underlying 
turbulence, such as the $ \alpha$-effect (which depends 
on the turbulent helicity) and turbulent diffusivity 
which then determine the mean field evolution 
\citep[see][for the details of formulation]{radler2014}.  
It is important to determine these turbulent transport 
coefficients to both compare with theoretical 
expectations and understand the working of the dynamo. 
This will be our aim here.

A number of different methods have been formulated and 
implemented so far to extract the dynamo coefficients. 
\citet{CH96} calculated the random magnetic field 
generated when a uniform field $\mean{ \mathbf{B}}$ is 
imposed in a helical turbulent flow, used it to find
$\mean{\mathbf{\mathcal{E}}}=\mean{{\mathbf{u}\times
\mathbf{b}}}$ and inverted the relation $\mean{\cal 
E}_i=\alpha_{i\!j} \mean{ B }_j$ to estimate the 
turbulent coefficients $\alpha_{i\!j}$. \citet{angstrom} 
adapted an experimental method developed by {\AA}ngstrom 
for measuring the conductivity of solids, to determine 
the large-scale diffusivity of magnetic fields, in two 
dimensional systems.

In another approach which can also handle additive noise,
\citet{BranSok02} (hereafter BS02) and 
\citet{Kowal06} computed different moments of mean fields 
with themselves and the EMF and fitted their linear 
relation with the data to extract the dynamo coefficients. 
More sophisticated methods like the test-field method 
({TF}) have also been previously used to estimate 
dynamo coefficients in direct numerical simulations of 
forced helical turbulence, ISM turbulence driven by 
supernovae, accretion disk turbulence, and convective 
turbulence in the context of Solar and Geo-dynamos 
\citep{schriner_test,schriner_test1,Bran05,sur2007kinetic,gressel_2008,kapala_test,bendre2015dynamo,GP15,War17}. This 
method relies on the idea that the fluctuating velocity 
$\mathbf{u}$ determined from solving the 
magnetohydrodynamic equations in any turbulence 
simulation contains all the information about the 
turbulent transport coefficients, in both the kinematic 
and dynamic phases. One then solves for the small scale 
magnetic field $\mathbf{b}_T$ induced by $\mathbf{u}$ 
acting on additional passive large-scale test fields 
$\mean{\mathbf{B}}_T$ with well defined functional forms, 
along with the direct simulations. The relation of the 
associated additional components turbulent EMF $\mean{
\mathbf{\mathcal{E}}}_T=\mean{ \mathbf{u}\times \mathbf{b
}_T}$ to $\mean{\mathbf{B}}_T$ is then used to determine 
the underlying dynamo coefficients 
\cite[See][for an overview and more remarks on TF method]{Brandenburg2009,brandenburg_2018}.

An alternative direct approach has been implemented by 
\citet{racine2011mode} and \citet{simard2016characterisation}, 
wherein the computation of dynamo coefficients is handled 
as a problem of least-square minimisation. Specifically, 
the time series of the EMF is fitted as a linear function 
of the mean field and mean current time series using the 
singular value decomposition (SVD) method. One convenience 
of this method over the {TF} is that it could be 
used as a post processing tool for the simulation data 
thereby making it computationally less expensive. In 
contrast to the {TF} method, where one solely uses 
$\mathbf{u}$  from DNS, one here additionally uses the 
actual $\mathbf{b}$ obtained directly from the simulation 
to calculate $\mean{\mathbf{\mathcal{E}}}$ and fit its 
relation to $\mean{\mathbf{B}}$ also obtained from the 
simulation, to estimate the dynamo coefficients. Thus 
there is no ambiguity in the applicability of the SVD 
method, at least in regimes where the transport 
coefficients can be assumed to be constant -- that is, 
both in the kinematic and fully-quenched regimes of the 
dynamo.

In view of these possible advantages, we explore here the 
SVD method as a tool to recover the turbulent transport 
coefficients in the previously published galactic dynamo 
simulation of \citet*{bendre2015dynamo}. These were 
magnetohydrodynamic simulations of a local box of 
stratified ISM, with turbulence driven via SN explosions. 
Specific parameters in the simulation domain were set 
to partially mimic the conditions in a galaxy like the 
Milky Way. In these simulations, it was found that 
large-scale magnetic fields emerge with an e-folding 
time of about 200\Myr. We analyze specifically the time 
series data from one of these runs to estimate the values 
of the turbulent transport coefficients using the SVD 
method. An added advantage is that we can also compare 
the results obtained here using the SVD method with the 
results obtained earlier for the same run from the 
{TF} method.

The paper is structured as follows. In 
\sref{sec:nirvana_setup} we describe the numerical setup 
for the direct numerical simulations (DNS) of the 
turbulent interstellar medium (ISM), followed by a brief 
discussion of its results in \sref{subsec:dns_results}. 
In \sref{sec:dynamo} we summarize the mean-field 
formulation and the algorithm we adopt for the extraction 
of dynamo coefficients. {\aref{app_mockdata} tests 
the SVD algorithm on mock data.} Results of our SVD 
analysis are discussed in 
\sref{sec:dns_dynamo_coefficients} and 
\sref{sec:1-d_dynamo}. {A comparison of these results 
with that obtained previously with the {TF} method 
in the kinematic phase is given in \aref{alphas_tf}. 
\aref{sec:regr} compares the SVD results with that from 
a simple regression analysis using the method of {
BS02}.} The final section presents a discussion of our 
results and our conclusions. 

\section{Direct Numerical Simulations}
\label{sec:nirvana_setup}
We briefly recall the setup and results of the DNS of
galactic dynamo that is analyzed here. A detailed
description of the numerical setup is also presented
in \citet{bendre2015dynamo,Bendre2016}.

The NIRVANA code \citep{nirvana} was used to simulate
the multi-phase ISM in a local Cartesian box ($ L_x =
L_y = 0.8 $\kpc) of the Galaxy. To study the vertical
distribution of the turbulent properties; we use disc
of thickness $\sim 4 $ \kpc ($-2.12$\kpc $< z < 2.12$
\kpc). The simulations use $96\times96\times512$ grid
cells which gives a numerical resolution of $\sim8.3$ 
\pc. We impose the shearing periodic boundary conditions 
in the radial, $x$ direction to match the differential 
rotation, and periodic in the azimuthal, $y$ direction 
to account for the approximate axisymmetric azimuthal 
flows observed in the disc galaxies. The galactic 
rotation curve is taken to be flat, with angular 
velocity decreasing with radius $R$ as $\Omega \propto 
1/R$, and having a value $\Omega_0=100$ \kms\kpc$^{-1}$ 
at the centre of the simulation box. Furthermore, 
outflow conditions are used at the vertical boundaries 
to allow the outflow of gas from the boundaries while 
preventing inflow.

Turbulence is driven via SN explosions, the locations
of which are chosen randomly with a prescribed rate of
$\sim 7.5$ \kpc$^{-2}$\Myr$^{-1}$, almost a quarter of
the average SN rate of Milky Way. The SN explosions
are simulated as localized Gaussian expulsions of
thermal energy. A stratified vertical profile of ISM
mass density with a scale-height of $\sim300$\pc (and
midplane value of $10^{-24}$\g\pcc) is also set up as
the initial condition and it is initially in hydrostatic
equilibrium under gravity. The vertical profile of
gravity is adapted from 
\citet{gravity_1,gravity_2,gravity_3}. To further capture 
the multi-phase morphology of ISM we adopt an optically 
thin radiative cooling function as a piece wise power 
law, $\Lambda \left( T\right)=\Lambda_iT^{\beta_i}$. The 
cooling coefficients of different ISM thermal phases 
$\Lambda_i$ are chosen similar to 
\citet{radiative_cooling}. This prescription does not 
quite capture the detailed cooling processes in highly
dense cold environments, although the primary goal here
is to simulate the dynamical aspects of ISM at moderate
densities and large length scales. Our initial magnetic
field profile was chosen to have a net vertical flux,
and strength of  $\sim$\nG which is 3-4 orders of 
magnitude smaller than the equipartition strength. This 
numerical set-up corresponds to model `Q' from our 
previous analysis
\citep{gressel2012magnetic,bendre2015dynamo,Bendre2016}
and details of the various source and sink terms are 
included therein.

\subsection{Evolution of Mean Fields in the DNS}
\label{subsec:dns_results}
The magnetic field, $\mathbf{B}$, in this setup amplifies
exponentially during the initial kinematic phase with
e-folding time of $\sim200 $\Myr, until it grows to 
approximately equipartition strength within $ \sim1$\Gyr. 
After reaching near equipartition values, the magnetic 
field continues to grow exponentially. We refer the 
initial amplification phase of $\sim1$\Gyr as the 
kinematic and the later as the dynamical phase. To 
explore the behaviour of large and small-scale fields 
separately, we define the mean components of $\mathbf{B}$ 
and $\mathbf{U }$ by averaging them over the $x$-$y$ 
(i.e., radial-azimuthal) plane, so as to have only $ z$ 
as an independent variable. Thus we define
\begin{align}
    \overline{\mathbf{B}}\left(z,t\right)=\frac{1}{L_x L_y}\,\iint \mathbf{B}\left(x,y,z,t\right) \,dx\,dy ,\nonumber \\
    \overline{\mathbf{U}}\left(z,t\right)=\frac{1}{L_x L_y}\,\iint \mathbf{U}\left(x,y,z,t\right) \,dx\,dy.
    \label{avg}
\end{align}
This definition of averaging in the current setup satisfies
the Reynolds averaging rules. Moreover, the $z$ component
of $\overline{\mathbf{B}}$ stays unchanged throughout the
evolution; subject to the solenoidality constraint. Also 
the $x$ component of mean velocity stays negligibly small
compared to the $z$ component - the outward wind, which 
has a linear profile in the $z$ direction. Both 
$\overline{\mathbf{B}}$ and the turbulent field, 
${\mathbf{b}}$, in the DNS have the same growth rate of 
$\sim200$\Myr during the kinematic phase, which later 
slows down in the dynamical phase, identical to the 
behaviour of the total magnetic field 
\citep[see Figure 2. from][]{gressel2012magnetic}.
Further, both $\overline{\mathbf{B}}$ and ${\mathbf{b}}$
have an approximate bell-shaped vertical profiles that
peak at the midplane 
\citep[see Fig. 3.3 and 3.4 from][]{Bendre2016}. The scale 
height and peak strength of mean field are approximately 
$0.6$\kpc  and $3$\muG,  respectively at $t=2.5$ Gyr, 
i.e., the end of the simulation. The growth of the mean 
magnetic field energy density is shown below in the right 
panel of Fig.~\ref{comparison_1}. Further, the space-time 
diagram of $\mean{B}_x$ and $\mean{B}_y$ are shown in the 
bottom panels of \fref{comparison_2}. These will be 
discussed further below while comparing with the SVD 
predictions.

\section{The Mean field dynamo}
\label{sec:dynamo}

The amplification of the large-scale magnetic fields is
generally understood using the mean-field dynamo theory
\citep{Mof78}. In mean-field theory, the magnetic
$\mathbf{B}$ and velocity $\mathbf{U}$ fields are separated
into their corresponding large and small scale components.
In particular as described above we write $\mathbf{ B}=
\overline{\mathbf{B }}+ \mathbf{ b }  $ and $\mathbf{U} =
\overline{\mathbf{U}} +  \mathbf{ u } $, where the average
is calculated over a suitable domain (in our case, over
$x$-$y$ plane as defined in \eref{avg}). The evolution of 
mean magnetic field is then governed by the averaged 
induction equation,
\begin{align}
\frac{\partial \overline{\mathbf{B}}}{\partial t} =
\nabla \times \left(
\overline{\mathbf{U}} \times \overline{\mathbf{B}} + \mean{\mathbf{\mathcal{E}}} -
\eta_m \nabla \times \overline{\mathbf{B}}
\right)
\label{mfe}
\end{align}
where $\mean{\mathbf{\mathcal{E}}}=\mean{\mathbf{u}\times
\mathbf{b}}$ is the turbulent EMF and $\eta_m$ the 
microscopic diffusivity.

Using the well established Second-Order Correlation 
Approximation \citep[SOCA, ][]{Mof78,radler2014}, the 
turbulent EMF can be expanded in terms of the mean 
field and its gradient as
\begin{equation} 
\mean{\mathcal{ E}}_i =\alpha_{i\!j} \mean{B}_j-
\eta_{i\!j}(\nabla\times\mean{\mathbf{B}})_j\,.
\end{equation}
For brevity of notation, in what follows, we set $\nabla
\times\mean{\mathbf{B}}=\overline{\mathbf{J}}$ the mean 
current density adopting $\mu_0 =1$. Dynamo coefficients 
$\alpha_{i\!j}\left(z,t\right)$ and $ \eta_{i\!j}\left(z
, t\right) $ are the tensorial quantities that depend on 
the properties of background turbulence. More explicitly, 
the turbulent EMF for this numerical setup is written as,
\begin{align}
 \begin{pmatrix} \,\, \overline{\mathcal{E}}_{x} \,\,\\ \overline{\mathcal{E}}_{y} \end{pmatrix}
 =
  \begin{pmatrix}
   \alpha_{xx} & \alpha_{xy}\\
   \alpha_{yx} & \alpha_{yy}
   \end{pmatrix}\,
\begin{pmatrix}
  \,\,\, \overline{{B}}_x \,\,\\ \,\, \overline{{B}}_y
   \end{pmatrix}
   -
  \begin{pmatrix}
   \eta_{xx} & \eta_{xy} \\
   \eta_{yx} & \eta_{yy}
   \end{pmatrix}\,
\begin{pmatrix}
  \,\,\, \overline{{J}}_x \,\,\\ \overline{{J}}_y
   \end{pmatrix}
   \label{e1}
\end{align}
Diagonal elements of the $\alpha$ tensor represent the
alpha-effect, proportional to the kinetic helicity
of the turbulence when the magnetic field is not
dynamically important and the turbulence is isotropic.
The anti-symmetric part of the off-diagonal components, 
$\alpha_{xy}$ and $\alpha_{yx}$, can be combined to 
produce the so called gamma-effect, $\gamma= 0.5\, 
\left(\alpha_{yx} - \alpha_{xy}\right)$, sometimes 
called ``turbulent pumping''. It leads to a component
of $\mean{\emf} = \gamma \times \mean{\mathbf{B}}$ and 
so advects the mean magnetic field similar to the mean 
velocity $\mean{\mathbf{U}}$. The diagonal components 
of the $\eta$ tensor represent the turbulent 
diffusivity of the mean magnetic field by small-scale 
motions and the off-diagonal terms can lead to for 
example the \citet{Radler69} effect.

\subsection{Determination of Dynamo Coefficients}
\label{sec:coefficients}
In order to invert \eref{e1} and compute all eight
dynamo coefficients, one needs a sufficient number 
of independent data points. In our previous work 
\citep*{bendre2015dynamo}, we used the {TF} 
method to measure these coefficients.

The current analysis, in contrast, relies only upon 
the simulation data, and uses the SVD method to perform 
least-square fit of mean field and mean-current data 
to the EMF data to extract the dynamo coefficients. 
This is similar to the method used by 
\citet{simard2016characterisation,racine2011mode}, for 
the analysis of thermally driven convective turbulence
in solar MHD simulations. We now turn to the detailed
implementation of SVD in the present setting.

\subsection{The Singular Value Decomposition Method}
\label{sec:svd_method}
The SVD method relies only upon the information of 
turbulent EMF and mean fields generated from the DNS. 
Here we compute the vertical profiles of 
$\mean{\mathcal{E}}\left(z,t\right)$, $\mean{\mathbf{B
}}\left(z,t \right)$ and $ \mean{\mathbf{ J}}\left(z,t
\right ) $ at various times, by averaging the DNS data
over $x$-$y$ plane and treat the time series as the data.
The extraction of the $\alpha_{i\!j}$ and $\eta_{i\!j}$ 
tensors is then achieved by fitting this data to the 
model described by \eref{e1} by minimising the square
of the residual vector components,
\begin{align}
        R_{i} =\mean{\mathcal{E}}_{i}
        - \alpha_{i\!j} \, \mean{B}_{j}
        + \eta_{i\!j}   \, \mean{J}_{j}
\label{residual_vector}
\end{align}
using the following algorithm.

Specifically, we extract the time series of the 
different components of turbulent EMF ($ \mean{
\mathcal{E}}_x$ and $ \mean{\mathcal{E}}_y  $), 
components of mean field ($\mean{B}_x$ and $\mean{B}_y$) 
and that of mean current ($\mean{J}_x$ and $ \mean{J}_y
$) at given $z=z'$ at independent times. If $ N$ is the 
length of these extracted time series $ \left( t_1,\,t_2 
, ... ,\,t_N\right) $, a design matrix $\mathcal{A}$ is 
defined as follows,
\begin{align}
\mathcal{A} =      \begin{pmatrix}
   \overline{{B}}_x\left(t_1,z'\right)
                &\overline{{B}}_y\left(t_1,z'\right)
                &-\overline{{J}}_x\left(t_1,z'\right) &-\overline{{J}}_y\left(t_1,z'\right)   \\
   \overline{{B}}_x\left(t_2,z'\right)
                & \overline{{B}}_y\left(t_2,z'\right)
                &-\overline{{J}}_x\left(t_2,z'\right) &-\overline{{J}}_y\left(t_2,z'\right)   \\
   \vdots
                & \vdots
                &\vdots
                &\vdots                                 \\
   \overline{{B}}_x\left(t_N,z'\right)
                & \overline{{B}}_y\left(t_N,z'\right)
                &-\overline{{J}}_x\left(t_N,z'\right) &-\overline{{J}}_y\left(t_N,z'\right)   \\
   \end{pmatrix}\,
   \label{design_matrix}
\end{align}
We note that the time series of mean field components 
and mean-current components form the different columns 
of $\mathcal{A} $, and each row corresponds to the values 
at any particular time, which makes $\mathcal{A}$, a 
matrix of dimensions $N\times4$. Since each of these 
columns are functions of $ z $, the matrix $\mathcal{A}$ 
also has a $ z $ dependence. This definition is used to 
write the following set of equations at $z=z'$ motivated 
by the model for EMF given in \eref{e1},

\begin{align}
    \mathcal{Y}\left(z'\right) = \mathcal{A}\left(z'\right)\, \mathcal{X}\left(z'\right) + \mathcal{N}\left(z'\right)
    \label{svd_equation}
\end{align}
where,
\begin{align}
\mathcal{Y}\left(z'\right) = \begin{pmatrix}
   \overline{\mathcal{E}}_x\left(t_1,z'\right)
        &\overline{\mathcal{E}}_y\left(t_1,z'\right)\\
   \overline{\mathcal{E}}_x\left(t_2,z'\right)
        &\overline{\mathcal{E}}_y\left(t_1,z'\right)\\
   \vdots       & \vdots                            \\
   \overline{\mathcal{E}}_x\left(t_N,z'\right)
        &\overline{\mathcal{E}}_y\left(t_1,z'\right)\\
   \end{pmatrix}\,
   \label{data_matrix}
\end{align}
\begin{align}
\mathcal{X}\left(z'\right)  =
  \begin{pmatrix}
   \alpha_{xx}\left(z'\right)   & \alpha_{yx}\left(z'\right)\\
   \alpha_{xy}\left(z'\right)   & \alpha_{yy}\left(z'\right)\\
     \eta_{xx}\left(z'\right)   & \eta_{yx}\left(z'\right)\\
     \eta_{xy}\left(z'\right)   & \eta_{yy}\left(z'\right)\\
   \end{pmatrix}\,
   \label{coefficients_matrix}
\end{align}
and the matrix $\mathcal{N}$ is the noise matrix, the
rows of which represent the level of noise in the data
at different times and is also a function of $z$. Note
that \eref{svd_equation} is completely consistent with
the SOCA model for EMF \eref{e1}, at each time. It
generalizes \eref{e1} to include a ``noise" component to
the EMF independent of mean field, which we can infer
from the SVD algorithm. We also assume that the matrix
$\mathcal{X}$, comprising of the dynamo coefficients
are time independent. This is expected to hold in the
kinematic regime (or during any period when the 
mean field grows exponentially) and the steady state 
saturation, but not during the transition between growth 
and saturation. Matrix $\mathcal{Y}$ and $\mathcal{X}$ 
are of dimensions $N\times2$ and $4\times2$ respectively,
while $\mathcal{N}$ has the same dimensions as $\mathcal{
Y}$.

For the $i^{\rm th}$ component ($i\in x,y$), 
$\overline{\mathcal{E}}_i$, of the turbulent EMF, at a 
given height $z =z'$, the data vector $\mathbf{y}_i
\left(z',t\right)$ is defined simply as the $i^{\rm th}$ 
column (i.e., the $1^{\rm st}$ and $2^{\rm nd}$ column 
for the $x$ and $y$ component, respectively) of the data 
matrix $\mathcal{Y}\left(z'\right)$, that is,
\begin{align}
    \mathbf{y}_i\left(z'\right)& =
    \begin{pmatrix}
      \overline{\mathcal{E}}_i \left(z',t_1\right)    \\
      \overline{\mathcal{E}}_i \left(z',t_2\right)    \\
      \vdots                                        \\
      \overline{\mathcal{E}}_i \left(z',t_N\right) \\
    \end{pmatrix}\,.
    \label{data_vector}
\end{align}
This data vector, $\mathbf{y}_i$, is also related to the
coefficient vectors $\mathbf{ x}_i$ (or the $i^{\rm th}$
column of matrix $\mathcal{X}$ in 
\eref{coefficients_matrix}). With these definitions 
\eref{svd_equation} can be rewritten separately for each 
column of $\mathcal{Y}$ as,
\begin{align}
        {\mathbf{y}_i}\left(z'\right)& = \mathcal{A}\left(z'\right) \, \mathbf{x}_i\left(z'\right) + \Hat{\mathbf{n}}_i\left(z'\right)\,,
    \label{e3}
\end{align}
where the vector $\Hat{ \mathbf{n}}_i $ represents the $
i^{\rm th}$ column vector of matrix $\mathcal{N}$. With 
this representation of EMF, the problem of estimation of
dynamo coefficients is the one of determination of
vector $\mathbf{x}_i$ that satisfies \eref{e3} at each $
z'$ separately. We note that \eref{e3}, comprises of $N$ 
simultaneous equations in four unknowns. The least square 
solution $\Hat{\mathbf{ x } }\left( z' \right)$ is the one 
that minimizes the two norm
\begin{align}
    \chi_i^2 \left(z'\right)&= \frac{1}{N}\,\displaystyle\sum_{n=1}^{N}\, \left[ \frac{\mathbf{y}_i(z', t_n) - \mathcal{A}\left(z',t_n\right) \, \mathbf{x}_i^\top\left(z'\right)}{\sigma_i} \right]^2
\label{e5}
\end{align}
at each height $z'$, and for each component $i$ (which 
can either be $x$ or $y$ in our case). Moreover $
\sigma_i$ is the variance associated with the noise 
matrix $\Hat{\mathbf{n}}_i$, which we assume 
independent of $n$ and will estimate post-facto from 
the fit itself (see below). The least square solution is 
obtained by employing SVD, which relies upon the unique 
decomposition of matrix $\mathcal{A}$ in the form,
\begin{align}
\mathcal{A}& = \mathbf{U}\, \mathbf{w}\, \mathbf{V}^\top\,,
\label{e6}
\end{align}
where $\mathbf{U}$ and $\mathbf{ V }$ are orthonormal
matrices and the matrix $\mathbf{w}$ is diagonal. One 
advantage of the representation of $\mathcal{A}$ as in 
\eref{e6} is that the least square solution vector 
$\Hat{\mathbf{x}}_i$ (components of dynamo coefficient 
tensors) is given simply by ``pseudo-inverting'' 
$\mathcal{A}$ \citep{mendel_svd,recepies} to yield
\begin{align}
\Hat{\mathbf{x}}_i = \mathbf{V}\, \mathbf{w^{-1}} \mathbf{U}^\top \mathbf{y}_i\,.
\label{e7}
\end{align}
The hat notation in the above equation is used to 
denote the least-square solution of \eref{e3} (which is 
different from $\mathbf{x}_{i}$ in \eref{e3}). The SVD 
method also gives an estimate for the covariance 
between the components of $\Hat{\mathbf{x}}_i$, in terms 
of the matrices $\mathbf{V}$ and $\mathbf{w}$. The 
covariance between $l^{\rm th}$ and $m^{\rm th}$ element 
of the vector $\Hat{\mathbf{x}}_j$ is given by
\begin{align}
        \mathrm{Cov}\Big([\Hat{\mathbf{x}}_j]_l, [\Hat{\mathbf{x}}_j]_m\Big) =  \sum_{i} \frac{\mathbf{V}_{li} \mathbf{V}_{mi}}{ \mathbf{w}_{ii}^{2}}\,.
\label{cov}
\end{align}
Note that the elements of the covariance matrix for 
both $\Hat{ \mathbf{ x}}_i$ ($i=1$ or $ i=2$) are the 
same, since the associated design matrix, $\mathcal{A}
$, is the same for both of them. Therefore the 
covariance between the identically indexed pairs of 
components of $\Hat{\mathbf{x}}_1$ and $\Hat{\mathbf{
x}}_2$ are the same, that is,
\begin{eqnarray}
\mathrm{Cov}\left(\alpha_{xx}, \alpha_{xy}  \right) & = &
 \mathrm{Cov}\left(\alpha_{yx}, \alpha_{yy}  \right)\,,\nonumber\\
 \mathrm{Cov}\left(\alpha_{xx},   \eta_{xx}  \right) & = &
 \mathrm{Cov}\left(\alpha_{yx},   \eta_{yx}  \right)\,,
\end{eqnarray}
and so on. The diagonal elements of \eref{cov} further 
provide a measure of the variance in the determination 
of individual fitting parameters. For instance, the 
error in the estimation of the $l^{\rm th}$ component 
of each $\Hat{\mathbf{x}}_i$ vector is given by,
\begin{align}
        \mathrm{Var}\Big([{\Hat{\mathbf{x}}_i}]_l\Big) &= \sum_{k} \left[\frac{\mathbf{V}_{lk}}{\mathbf{w}_{kk}}\right]^2\sigma_i^2\,.
\label{var}
\end{align}
The term in the square brackets is the $l^{\rm th}$ 
diagonal element of the covariance matrix defined in 
\eref{cov}. Here $\sigma_i^2$ is determined from the 
data, and the fitted parameter vector$\Hat{\mathbf{x
}}_i$, that is
\begin{align}
\sigma_i^2 = \frac{1}{N}\left(\mathbf{y}_i - \mathcal{A}\Hat{\mathbf{x}}_i\right)\left(\mathbf{y}_i - \mathcal{A}\Hat{\mathbf{x}}_i\right)^\top.
\end{align}
We note that the term in square bracket is the same for
both columns $\Hat{\mathbf{x}}_i$ (with $i\in 1,2$) but
$\sigma_i$ is different for the two.

In \aref{app_mockdata}, we have tested the SVD algorithm 
on noisy mock data generated by assuming dynamo 
coefficient profiles very similar to those we recover in 
the real data, running a 1-D mean-field dynamo model, 
and adding noise. We find the SVD method to be quite 
robust in recovering input parameters, if independent 
noise is added to the field and the current. If we add 
noise only to the magnetic field and derive the current 
from this, we find SVD method still recovers with good 
accuracy the $\alpha_{i\!j}$ tensor and with less 
accuracy the $\eta_{i\!j}$ tensor, the latter effect 
arising due to extra noise introduced in calculating the 
current from the field.

\section{Results of the SVD Reconstruction}
\label{sec:dns_dynamo_coefficients}

We use the 3-D DNS data for model Q, described in
\sref{sec:nirvana_setup} and construct the vertical 
profiles of $ \mean{\emf} = \mean{ \mathbf{u} \times 
\mathbf{b}}$, $\mean{\mathbf{B}}$ and $\mean{\mathbf{
J}}$, at each time step using \eref{avg}. We further 
smooth the profiles by applying a box filter with the 
window size equal to the SNR scale (which is 
approximately equivalent to the size of 4 grid cells, 
and this essentially filters out all noise below 
turbulence forcing scale). We note that this smoothing 
preserves the Reynolds rules. We then choose a time 
period corresponding to the range $0.1-1$\Gyr in the 
kinematic phase of the DNS model, and extract the time 
series of $\mean{\emf}$, $\mean{\mathbf{B}}$ and 
$\mean{\mathbf{J}}$, independently at each $z$. The time
interval between each data point is smaller than the 
expected correlation time of $10$ Myr 
\citep{anvar_2004,Bendre2016}. Thus we choose subsets 
of the full time series where data points are more than 
$10$ Myr apart such that each data point in the time 
series can be considered as independent. We construct 9 
such time series, referred to as $S1,S2,...,S9$, by 
starting from different initial times. For comparison, 
we also carry out the SVD analysis of the full time 
series, which we refer to as $S$.

From any of these time series as columns, the data 
matrices $\mathcal{Y}$ and design matrices 
$\mathcal{A}$ are constructed using definitions 
\eref{data_matrix} and \eref{design_matrix} 
respectively (i.e., separately at each $z$). Since 
the chosen time period corresponds to the kinematic 
phase of magnetic field we expect the constancy of 
dynamo coefficients in this period, i.e.  the time 
independence of coefficient matrix $\mathcal{X}$
(\eref{coefficients_matrix}). The mean field, current 
and EMF components from the DNS all grow exponentially 
and increase by almost three orders at the end of 
kinematic phase compared to their initial value. To 
compensate for this large growth, we multiply the time 
series of all mean field, current and EMF components by 
$\exp{\left(-t/200{\rm \Myr}\right)}$ and take out the 
exponential growth factor before fitting the data. 
{Choice of this particular scaling factor is based 
on the approximate exponential growth factor of mean 
magnetic energy, seen in the actual DNS data. We find 
the evolution of mean magnetic energy to be roughly 
proportional to $\exp{\left(t/100{\rm \Myr}\right)}$ 
(see the right panel of \fref{comparison_1}), a 
square-root of this factor is therefore used to 
approximate the exponential amplification of mean field 
and EMF components. We have checked that the results 
are insensitive to the exact choice of the scaling 
factor}. We apply the SVD algorithm described in 
\sref{sec:svd_method} to this data and determine the 
vertical profiles and variances of $\alpha_{i\!j}(z)$ 
and $\eta_{i\!j}(z)$ using \eref{e7}. In particular, we 
use the `svdcmp' algorithm described in \citet{recepies} 
to decompose the design matrix $\mathcal{A}$.

\begin{figure}
\centering\includegraphics[width=\columnwidth]{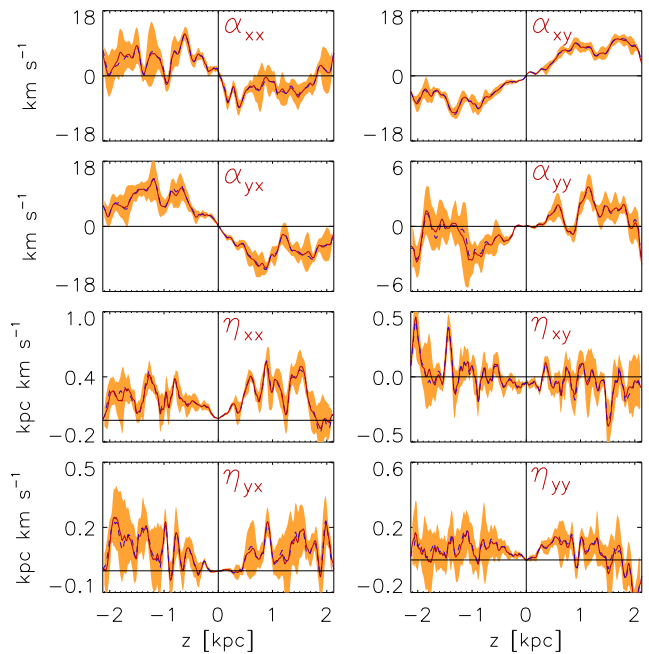}
        \caption{The red solid red lines show the average
        of the vertical profiles of different dynamo
        coefficients computed by applying SVD method to
        nine time different series $S1, S2,...,S9$ during
        the initial kinematic phase. Orange regions show
        the $1-\sigma$ variances obtained from these nine
        different vertical profiles. The inherent error
        in the SVD estimate of individual coefficients
        (from \eref{var}) are almost an order of magnitude
        smaller than the orange regions in the graph.
        The dashed blue line indicate the vertical
        profiles of the corresponding coefficients, calculated
        using the entire time series $S$ in the kinematic
        phase. We point out that the red solid lines and blue
        dashed lines almost coincide for all coefficients.
        Note that $\alpha_{xy}\approx-\alpha_{yx}$ giving
        rise to a vertical pumping term.}
\label{alpha_eta}
\end{figure}

The profiles of all the components of $\alpha_{i\!j}$ 
and $\eta_{i\!j} $ tensors, recovered using the SVD 
method, from the various time series are shown in 
\fref{alpha_eta}. The solid line in each panel shows 
the average profile obtained by averaging the 
individual profiles recovered from the time series $S1$ 
to $S9$, while the dashed line shows dynamo 
coefficients obtained from the full time series $S$. We 
see a close correspondence between these two, showing
that the oversampling implicit in the full time series 
$S$ does not affect the results. {In plotting we 
smooth these profiles over a window of size $100$\pc 
which corresponds roughly to the turbulent correlation 
length scales.}
Also shown in \fref{alpha_eta} by the orange shaded 
regions are the variance obtained in the dynamo 
coefficients recovered using the time series $S1$ to 
$S9$. We note that the formal error from the SVD 
analysis on the $\alpha_{i\!j}$ and $\eta_{i\!j}$
coefficients obtained using Eq.~\ref{var} are much 
smaller than this variance.

We see from \fref{alpha_eta} that the overall shapes 
of the profiles of diagonal $\alpha$ components are 
linear in the inner disc of approximate $-0.8$\kpc 
to $0.8$\kpc, and they have the opposite signs above 
and below the midplane. The magnitude of $\alpha_{yy
}$, which is the crucial part of the $\alpha$-effect 
for the $\alpha-\Omega$ dynamo, is zero at the 
midplane (as expected) and rises with $z$ to attain 
a maximum of about $3 {\rm \km\s^{-1}} $ by $z=1$\kpc. 
A curious feature is that $\alpha_{xx}$ is larger and 
opposite in sign to $\alpha_{yy}$. Another significant
result of this analysis is the emergent antisymmetry 
of $\alpha$ tensor, which is to say that off-diagonal 
elements are of opposite signs and similar magnitude. 
As already found previously \citep{gressel_2008}, 
these two combine, to constitute a turbulent pumping 
term $\gamma\sim10$\kms at the height of $z=1$\kpc 
that acts to transport mean magnetic fields towards 
the equator, against the outward advection by the 
vertical velocity $\overline{U}_z$. The profiles of 
the $\eta$ tensor, as recovered by the SVD method, 
are expectedly much noisier. The shape of both 
diagonal components of $\eta$ tensor profiles is
approximately inverted bell shaped, with a maximum 
turbulent diffusivity of $\simeq 10^{26} {\rm \cm^2 
\s^{-1}}$ at a distance of a kpc from the disk 
midplane. These values compare favorably with 
theoretical expectations \citep{anvar_2004}. The off 
diagonal component $\eta_{xy}$ oscillates around zero, 
while curiously $\eta_{yx}$ is also bell shaped, 
positive and similar to $\eta_{yy}$.

\begin{figure}
\centering\includegraphics[width=\linewidth]{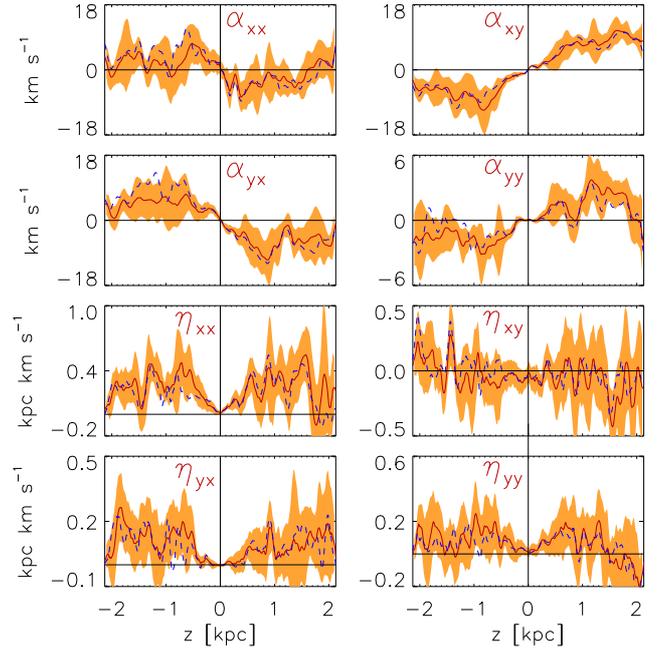}
\caption{Plotted in the red-solid lines are the average 
	vertical profiles of the dynamo coefficients 
	computed for four different sections of kinematic 
	phase (0.1 to 1 \Gyr) of size 225\Myr. Blue-dashed 
	lines represent the same for respective dynamo 
	coefficients, but for the entire kinematic phase 
	time series, these are identical to the ones 
	shown in \fref{alpha_eta}. Notice the significant 
	agreement between the red-solid and the blue
        -dashed curves. Orange regions in each panel 
	represent the corresponding 1-$\sigma$ interval 
	computed using the profiles of dynamo coefficients 
	in the four sections of kinematic phase.}
\label{alpha_eta_slots}
\end{figure}

We have also checked the robustness of SVD results 
in an alternate manner, by inter-comparing the 
coefficients computed within different sub-intervals 
of the original time series. In particular, we divide
the kinematic phase time series (100 to 1000\Myr) of 
mean field, mean current and EMF components, into 
four sections of equal lengths and compute all dynamo 
coefficients corresponding to each of these sections, 
using the SVD method. These four profiles are then 
used to compute the average profile and the dispersion 
about the average. In \fref{alpha_eta_slots} we show 
as red-solid curves, the average vertical profiles of 
the dynamo coefficients, obtained from the four 
different sub-intervals of kinematic phase. This can 
be readily compared with blue-dashed curves which 
shows the same dynamo coefficients, computed for the 
entire kinematic phase time series (same as the 
profiles shown in \fref{alpha_eta}). We see reasonably
good agreement between the two curves which shows the 
robustness of the SVD method and the validity of the 
assumption that the dynamo coefficients are 
approximately constant during the kinematic phase.
Represented by the shaded orange region is the 1-$
\sigma$ interval for these four vertical profiles of 
dynamo coefficients. They also show that the 
fluctuation of $\alpha_{i\!j}$ and $\eta_{i \!j}$ with 
time is larger than the formal error given by the SVD 
analysis using the full time series.

These profiles of dynamo coefficients recovered via 
SVD are in qualitative agreement with those 
recovered from {TF} method for the same model 
\citep{bendre2015dynamo, Bendre2016}. For ready
reference we have summarized the {TF} results 
derived using the same DNS data in \aref{alphas_tf}. 
The magnitude of $\alpha_{yy}$, recovered by the SVD 
method, is within a $1-\sigma$ confidence interval 
of its {TF} counterpart. However the magnitude 
of $\alpha_{xx}$ from the SVD method is systematically
larger by a factor of about 3. Furthermore, a bell 
shaped profile of both diagonal components of 
$\eta_{i\!j}$ tensor is obtained in both the SVD and 
{TF} methods. The magnitudes of $\eta_{xx}$ 
and $\eta_{yy}$ however, are substantially smaller 
in SVD  reconstruction, compared to the {TF} 
results. One possible explanation for this discrepancy 
is that the SVD and {TF} methods, in fact, may 
be sampling very different length scales in the 
problem, and that the different values for $\eta$ 
simply reflect this aspect. We will discuss this 
issue further in Section~\ref{discussion}.

{In addition, in \sref{sec:regr} the results 
from the SVD analysis are also compared with that 
from a simple regression method due to {BS02}. 
The mean values of the coefficients obtained in 
this method are in agreement with that obtained 
using SVD. The standard deviations tend to be 
however somewhat larger. Moreover as we will see 
below, we also use the SVD to compute systematically 
the full covariance matrix on the coefficients.}

\subsection{Covariance of the Dynamo Coefficients}
In principle, the overdetermined system defined by 
\eref{svd_equation} has no unique set of solutions 
in a sense that the same EMF time series could be 
produced by the different sets of parameters. The
SVD algorithm provides a solution that is the best
approximation in a least-squared sense. There 
however could be a degeneracy in the determination 
which could be probed quantitatively by comparing 
the off-diagonal elements of covariance matrix to 
the diagonal ones. Since the time series of both 
$  x  $ and $ y $ components of EMF (columns of 
$\mathcal{Y}$) depend on the first and second 
column of $ \mathcal{X} $ through the same design 
matrix $\mathcal{A}$ covariance between the 
elements of $\Hat{\mathbf{x}}_1$ and 
$\Hat{\mathbf{x}}_2$ are identical (see 
\eref{cov}) and following relations hold:
\begin{align*}
\mathrm{Cov}\Big(\alpha_{xx},\alpha_{xy}\Big) = \mathrm{Cov}\Big(\alpha_{yx},\alpha_{yy}\Big)\,,\\
\mathrm{Cov}\Big(\alpha_{xx},\eta_{xx}\Big)   = \mathrm{Cov}\Big(\alpha_{yx},\eta_{yx}\Big)\,,  \\
\mathrm{Cov}\Big(\alpha_{xx},\eta_{xy}\Big)   = \mathrm{Cov}\Big(\alpha_{yx},\eta_{yy}\Big)\,,  \\
\mathrm{Cov}\Big(\alpha_{xy},\eta_{xx}\Big)   = \mathrm{Cov}\Big(\alpha_{yy},\eta_{yx}\Big)\,,  \\
\mathrm{Cov}\Big(\alpha_{xy},\eta_{xy}\Big)   = \mathrm{Cov}\Big(\alpha_{yy},\eta_{yy}\Big)\,,  \\
\mathrm{Cov}\Big(\eta_{xx},\eta_{xy}\Big)     = \mathrm{Cov}\Big(\eta_{yx},\eta_{yy}\Big)\,.
\end{align*}

\begin{figure}
\centering
\includegraphics[width=\linewidth]{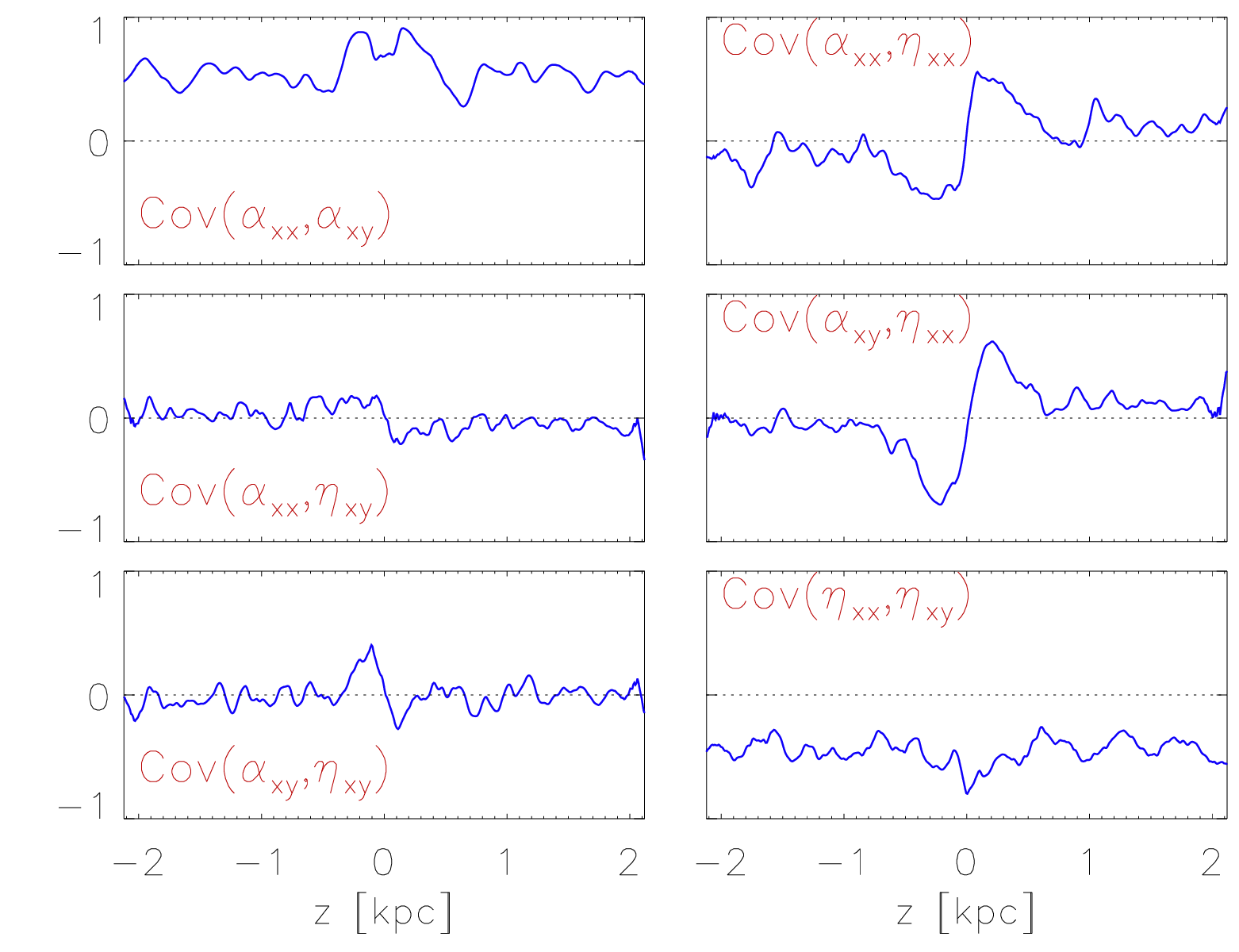}
        \caption{Vertical profiles of all off-diagonal 
	elements of covariance tensor normalized with 
	their respective diagonal components.}
\label{covar}
\end{figure}

In \fref{covar}, we plot the vertical profiles of 
the off-diagonal elements of the covariance matrix 
(calculated using \eref{cov}) normalized with the 
corresponding diagonal elements. If two parameters 
are uncorrelated the corresponding normalized 
covariance will be zero, full correlation 
corresponds to $+1$ and complete anti-correlation 
to $-1$. From this figure, we see that the diagonal
and off-diagonal elements of the $\alpha_{i\!j}$ 
tensor, like $\alpha_{xx}$ and $\alpha_{xy}$ (or 
$\alpha_{yy}$ and $\alpha_{yx}$) are correlated for 
all $z$, while the corresponding diagonal and 
off-diagonal elements of the $\eta_{i\!j}$ tensor 
are anti-correlated (see the top-left and 
bottom-right panels of \fref{covar}). Moreover, the 
correlations between the first and third element 
and the second and the fourth element of $\Hat{
\mathbf{x}}_i$ (equivalently $\mathrm{Cov}\left(
\alpha_{xx},\eta_{xx}\right)$ and $ \mathrm{Cov}
\left(\alpha_{yy},\eta_{yy}\right)$ are non 
negligible within the central half a kpc of the 
disk, where the mean field is strong.  

These correlations can arise if there are definite
correlations between the mean field components 
with themselves and with the current components. 
For example if we have a definite eigenmode of the 
mean-field dynamo with say $\mean{B}_y \approx - 
c_1\mean{B}_x$, with a constant $c_1$. Then 
$\mean{\mathcal{E}}_x = \alpha_{xx}\mean{B}_x + 
\alpha_{xy}\mean{B}_y+...\approx (\alpha_{xx} -c_1
\alpha_{xy}) \mean{B}_x + ....$, and there can be 
a mixing (or correlation) between $\alpha_{xx}$ 
and $\alpha_{xy}$, as indeed observed. Similarly, 
if the field is partially helical, then there would 
be a correlation between $\mean{B}_x$ and $\mean{J
}_x$ (also $\mean{B}_y$ with $\mean{J}_y$). This 
could indeed induce a partial correlation between 
$\alpha_{xx}$ and $\eta_{xx}$ (and $\alpha_{yy}$ 
with $\eta_{yy}$), as also indeed observed in the 
top right panel of \fref{covar}.

We see therefore that all the turbulent dynamo 
coefficients determined by the SVD, by directly 
fitting the $\mean{\mathbf{\mathcal{E}}}$ data
from the numerical simulation, need not be 
completely independent. This will be the case if 
the mean fields and currents are partially 
correlated. The coefficients determined by the 
SVD do however give the best fit to the data in 
a least-squared sense. This also explains why the 
{TF} results and the SVD method, even if 
they differ somewhat in the amplitudes of 
different coefficients, could lead to very 
similar $\mean{\mathbf{\mathcal{E}}}$ and hence 
predict similar evolution for the mean field 
evolution.

\subsection{Quenching of the Dynamo Coefficients}

\begin{figure}
\centering
\includegraphics[width=\columnwidth]{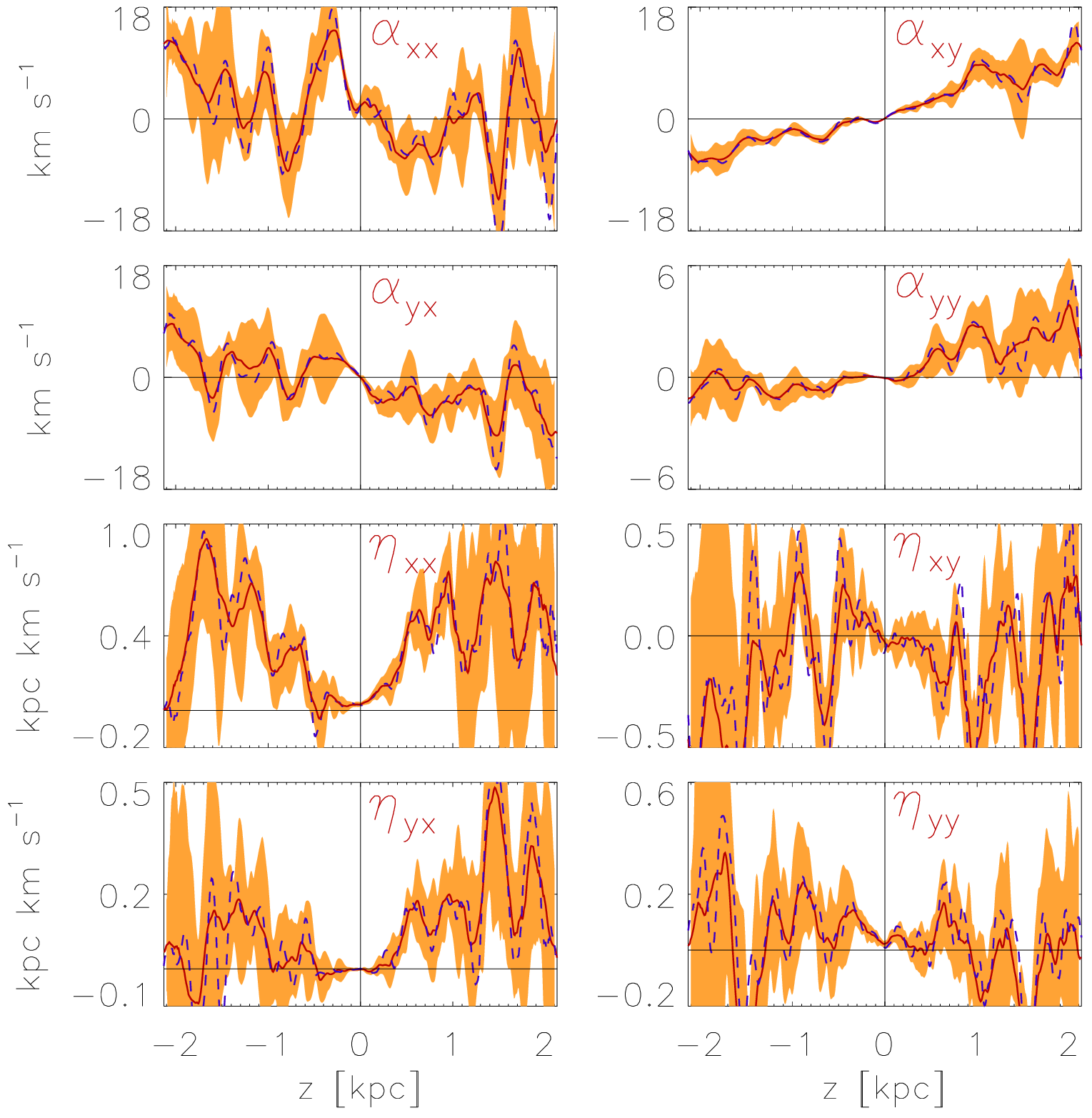}
\caption{Same as \fref{alpha_eta} but for the 
	dynamical phase.}
\label{alpha_quench_all}
\end{figure}

\begin{figure*}
\centering
\includegraphics[width=0.8\linewidth]{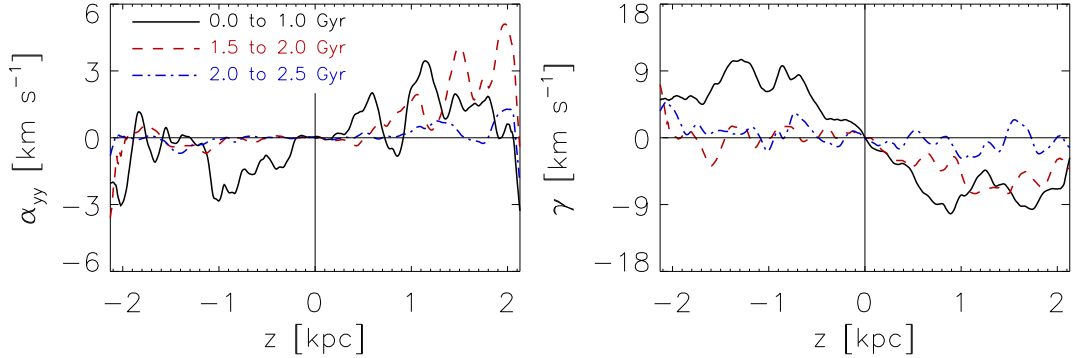}
\caption{\textit{Left Panel:} Vertical profile of 
	the $\alpha_{yy}$ coefficient in the 
	kinematic (black-solid line), the dynamical 
	phases ($1.5$ - $2.0$) Gyr (red-dashed line) 
	and ($2.0$ - $2.5$) Gyr (blue-dash-dotted 
	line). \textit{Right Panel:} shows the same 
	but for turbulent pumping (i.e., $\gamma$) 
	term with the same colour coding.}
\label{alpha_quench}
\end{figure*}

It is of interest to examine the behaviour of 
the dynamo coefficients as the mean magnetic 
field grows and Lorentz forces become important. 
Our previous analysis based on the {TF}
method \citep{gressel2012magnetic,Bendre2016} 
indicated that both $\alpha$ and $\eta$ 
coefficients quench drastically in the presence 
of dynamically significant mean fields as an 
algebraic function of relative field strengths. 
We perform a similar analysis using SVD method 
here. To quantify the strengths of the 
mean fields relative to turbulent kinetic 
energy we use the dimensionless ratio of the 
magnetic to kinetic energy defined by,
\begin{align}
        \beta^2 = \frac{\mean{\mathbf{B}}\cdot\mean{\mathbf{B}}}{\mu_0 \,\rho\, u^2}\,.
\end{align}
To investigate the effect of strong mean field 
on the dynamo coefficients, we examine their 
behaviour at two regions of $\beta$. We choose 
the time series of $\mean{ \mathbf{ B }}$,
$\mean{\mathbf{J}}$ and $\mean{\emf}$ from DNS, 
corresponding to $0.1$ to $1 $\Gyr representing 
its initial kinematic phase ($\beta\leq0.01$). 
Conversely, for the dynamical phase ($\beta\geq
1$) we use the time series between $1.5 $ \Gyr 
and $2.54 $\Gyr. Equipped with this data, we 
compute vertical profiles of all dynamo 
coefficients using the SVD method as discussed 
in the previous section.

It should be noted that, the SVD algorithm we 
have employed requires the dynamo coefficients 
to stay constant for a chosen range of time. 
Consequently while choosing  the time slots at
various values of $\beta$ we are tacitly 
assuming the constancy of dynamo coefficients 
for that range. This assumption may be 
justifiable in the kinematic phase where the 
exponential growth of magnetic is consistent 
with the solution of $\alpha-\Omega$ dynamo 
with constant dynamo coefficients. Moreover,
in the dynamical phase above $\sim 1.5$\Gyr, 
the average growth of mean fields appears to 
be approximately exponential, however with a 
drastically reduced growth rate. This also 
hints the existence of dynamo action with a 
set of approximately constant (but quenched) 
dynamo coefficients. For the intermediate 
phase of approximately 1.0 and 1.5\Gyr however, 
where the transition between kinematic and 
quenched dynamo coefficients is supposed to 
occur, the constraint of the constancy may not 
hold. We therefore focus here on the dynamo 
coefficients in the kinematic and dynamical
phase using this method, and not in the 
intermediate phase of transition.

The dynamo coefficients in the dynamical phase 
are shown in \fref{alpha_quench_all}. The 
plots show a drastic suppression of the mean 
$\alpha_{yy}$ and the off-diagonal terms, 
$\alpha_{xy}$ and $\alpha_{yx}$, for $ z < 0$, 
which in turn implies a drastic suppression of 
the pumping term $ \gamma=\left(\alpha_{ yx} - 
\alpha_{ xy }\right)/2$. It appears that the 
mean value of these coefficients are less 
affected for $z>0$.
However, the lower envelope of $\alpha_{yy}$ 
is closer to zero in the dynamic phase
compared to the kinematic phase, even for $z
>0$.
We find that the other coefficients do not 
undergo any systematic quenching, although the 
variance in these coefficients are much larger
in the dynamic compared to the kinematic phase 
{and it is harder therefore to 
quantitatively determine the quenching}.

To elucidate the behaviour of the dynamo 
coefficients in the dynamical phase further, we 
have split this period in to two sub-periods, $
1.5$ to $2$ \Gyr, and $2$ to $2.5$ Gyr. In 
\fref{alpha_quench} we compare the vertical 
profiles of $ \alpha_{yy}$ and $\gamma$ 
coefficients for these two dynamical regimes 
with the kinematic regime. The figures show
clearly, that for the final period of  $2$ to 
$2.5$ Gyr, the suppression of $\alpha_{yy} $ 
and the pumping term $\gamma$ is now drastic 
for all values of $z$, This also shows that the 
turbulent coefficients are evolving in the 
dynamic phase and \fref{alpha_quench_all} shows 
the average behaviour, assuming they are 
constant.

Note that in the kinematic stage the $\gamma$ 
term pumped back the mean field into the disk 
part, which an outflowing wind was carrying out 
in the halo. The decrease of $\alpha_{yy}$ and 
$\gamma$ could therefore be the cause of 
dynamical saturation of magnetic field seen in 
the DNS. We should also point out that the mean 
vertical velocity $\overline{U}_z$ is suppressed 
as the $\beta$ or the mean magnetic field 
increases, as already shown in 
\citet*{bendre2015dynamo}.

\subsection{Comparison of recovered \texorpdfstring{$\mean{\emf}$}{E} with the DNS}
\label{DNS_SVD_emf}

\begin{figure*}
\centering\includegraphics[width=0.9\linewidth]{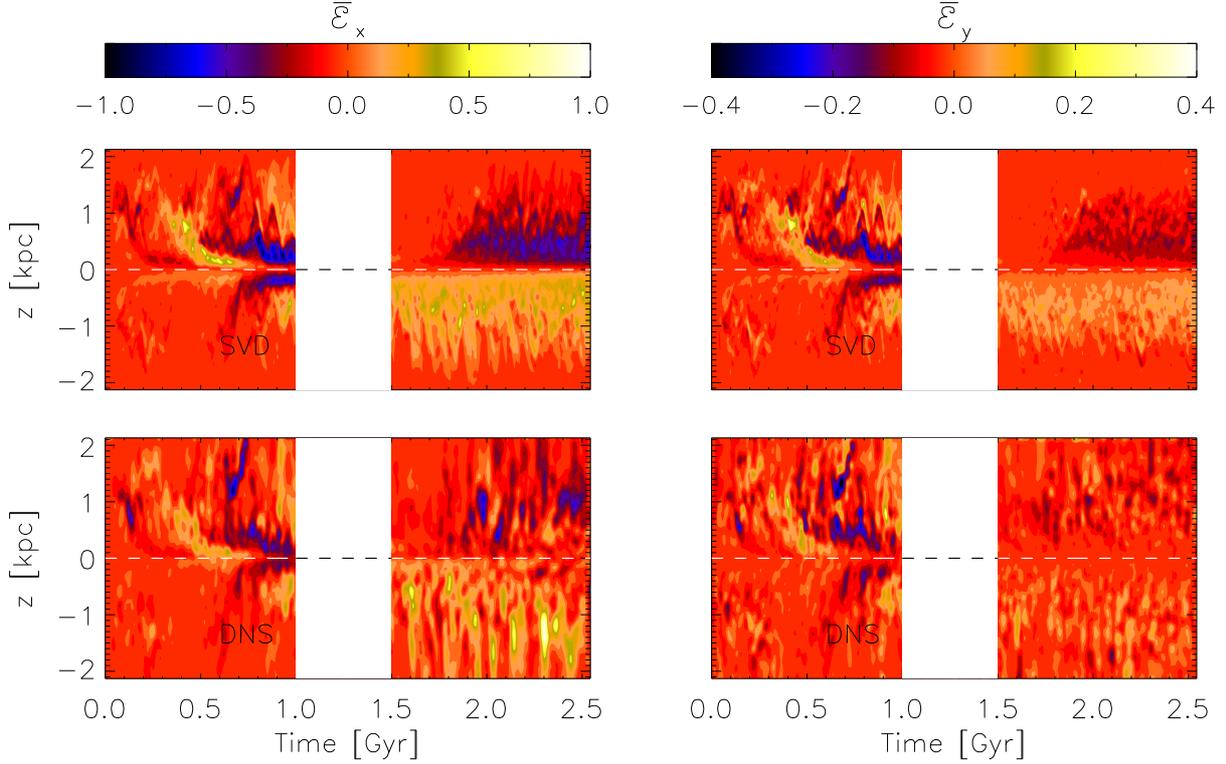}
\caption{\textit{Left Panels:} Bottom panel shows 
	the time evolution of the vertical profile 
	of $\overline{\mathcal{E}}_x$ obtained from 
	the DNS and the top panel depicts the same
        for the $x$ component of reconstructed EMF 
	using the SVD estimates of $\alpha$ and $
	\eta$ tensors and the profiles of mean field 
	components from DNS. \textit{Right Panels:} 
	Show the same quantities as the left one 
	but for the $y$ component of EMF. Note that 
	the SOCA the model for EMF components 
	roughly reproduces the actual DNS data. The 
	white patch ranging from 1 to $1.5 $\Gyr 
	corresponds to the transition phase between 
	the kinematic and dynamical phase of field
    evolution, since the dynamo coefficients in 
	this range cannot be reliably extracted by 
	present SVD method, we have omitted this 
	patch in this comparison. Color code here is normalized with respect to the exponential 
	scaling factor of $\exp{\left(t/200 {\rm Myr}
	\right)}$, in the kinematic phase to compensate 
	for the exponential amplification of EMF 
	components as mentioned in \sref{sec:svd_method}.}
\label{emf_contour}
\end{figure*}

It is important to ask how well $\mean{\emf}$, 
obtained from the turbulent transport 
coefficients recovered with the SVD method 
according to \eref{e1} or \eref{svd_equation} 
agrees with $\mean{\emf}$ in the DNS and what 
is the level of residual noise in this fit. 
We compare in \fref{emf_contour} the evolution 
of  $\overline{ \mathcal{E}}_x \left(z\right)$ 
and $\overline{\mathcal{E}}_y\left(z\right)$
obtained from the DNS (Bottom panels) with 
that reconstructed using the SVD estimates of 
the $\alpha_{i\!j}$ and $\eta_{i\!j}$ tensors 
(top panels). We see that that there is 
reasonable agreement between the two, 
especially in the kinematic stage.

\begin{figure}
\centering
\includegraphics[width=\linewidth]{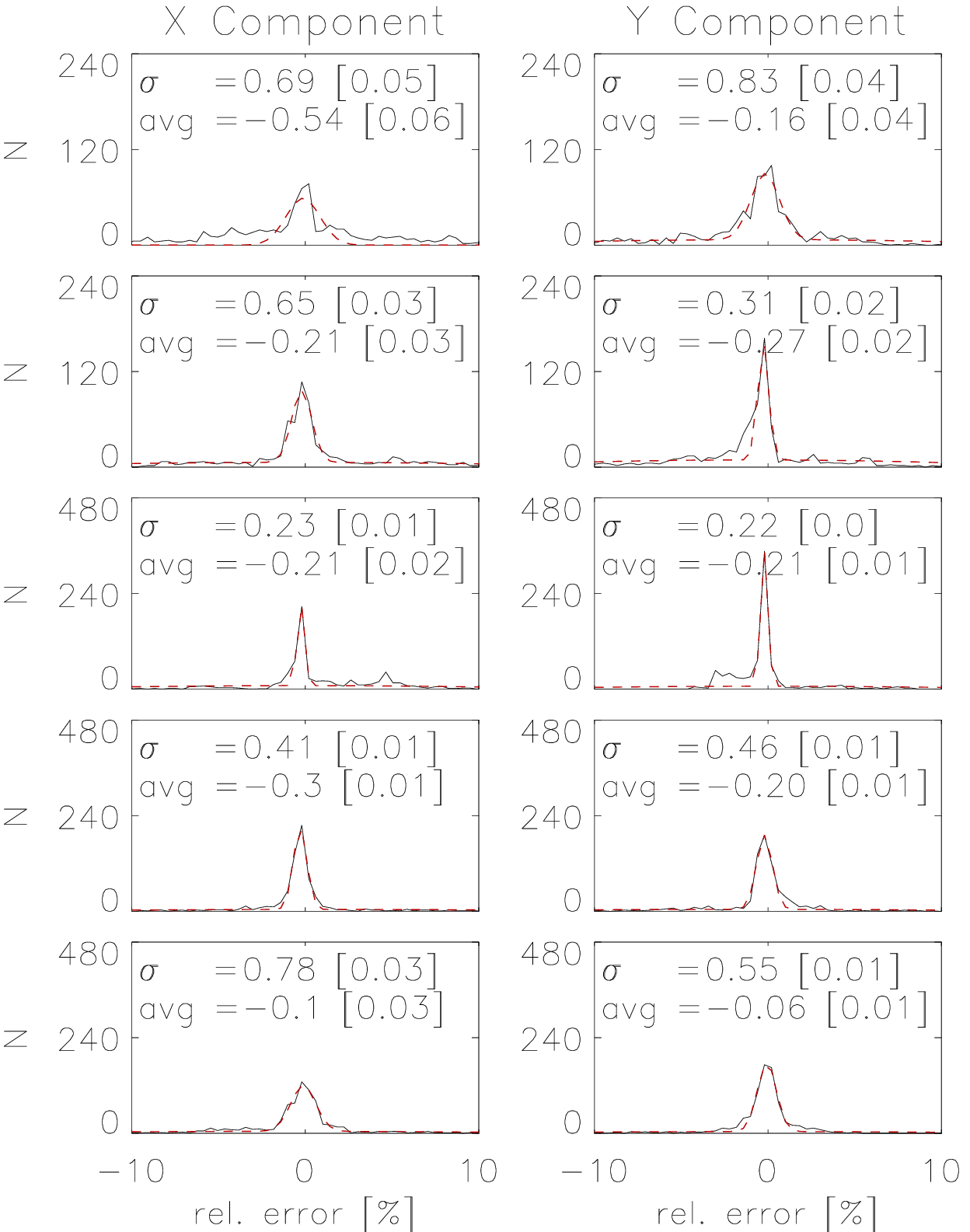}
\caption{Black-solid lines show the residual noise 
	`probability' distributions obtained by 
	subtracting the {kinematic phase} time 
	series of $\mean{\mathbf{\mathcal{E}}}$ 
	{obtained in the SVD reconstruction} 
	from {that} obtained in 
	the DNS. The noise is expressed in the 
	units of percentage of the DNS $\mean{
	\mathbf{\mathcal{E}}}$ value. The red lines 
	show the Gaussian best fit of the 
	respective distributions. Mean (avg) and 
	standard deviation ($\sigma$) of each fit 
	is given in each box {and the square 
	brackets show the errors in their 
	determination}.}
\label{emf_noise_kin}
\end{figure}

\begin{figure}
\centering
\includegraphics[width=\linewidth]{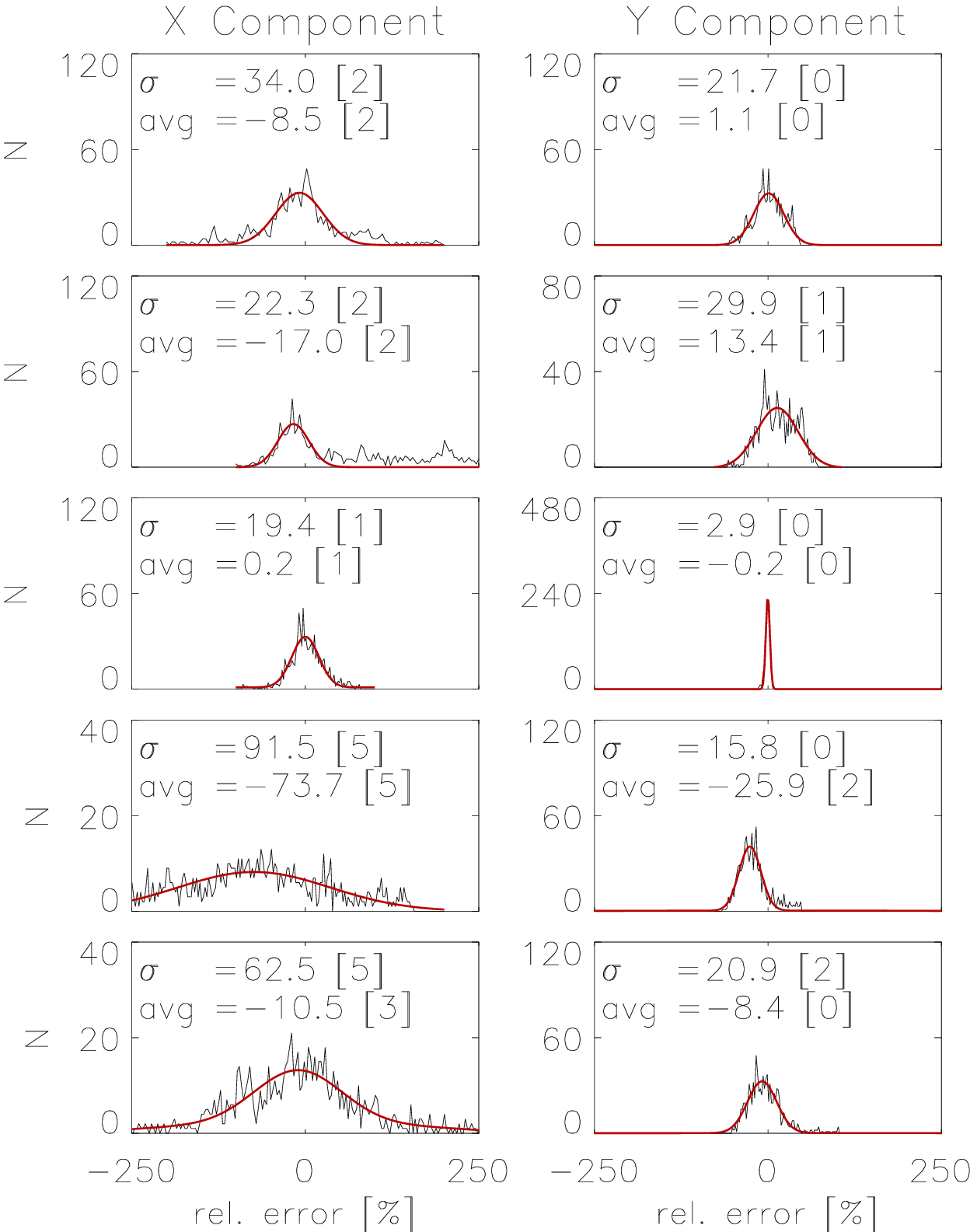}
\caption{Same as \fref{emf_noise_kin} but for the 
	dynamical phase time series (restricted to 
	the time period of $t \ge1.5$\Gyr) of EMF 
	components.}
\label{emf_noise_dyn}
\end{figure}

In order to make this comparison more quantitative 
and estimate the level of the residual noise in the
SVD fit, we have also compared the time-series of 
both components of EMF obtained from the DNS ($
\mean{\mathcal{E}}_i=\mean{\left(\mathbf{u}\times 
\mathbf{b}\right)_i}$) and the ones reconstructed 
using the dynamo coefficients ($\mean{\mathcal{E}
}_i = \alpha_{i\!j}\mean{B}_j-\eta_{i\!j}\mean{J
 }_j $) at specific locations. This is shown in
\fref{emf_noise_kin} and \fref{emf_noise_dyn}, 
where we plot the histograms of the relative 
differences (in percentage) between $x$ and $y$ 
components of EMF obtained from the DNS and 
corresponding estimates using SVD, {for the 
kinematic ($0.1$ to $1$\Gyr) and the dynamical 
phase (above $1.5$ \Gyr) respectively. The left 
and right panels of each figure, correspond 
respectively to the distribution of this residual 
noise in the $x$ and $y$ components of EMF. We do 
this analysis at various heights and the panels 
from bottom to top show the results at $z=-1$\kpc, 
$-0.5$ \kpc, $0$\kpc, $0.5$ and $1.0$ \kpc}. We 
see that the relative difference between the two 
is mostly normally distributed at all heights. We 
{then} fit these histograms with the Gaussian
functions {(shown with red curves)}, and 
determine both the mean and the dispersion $
\sigma$ of the distributions.

We see from \fref{emf_noise_kin} that the mean 
of the residual noise is very close to zero 
during the kinematic phase, while their $\sigma$ 
values turn out to be less than a few percent at 
all locations. Therefore the reconstructed EMF 
using the SVD method does give a good fit to that 
obtained directly in the simulations during the 
kinematic evolution. On the other hand, both the 
mean residual noise and its dispersion are larger 
during the dynamical phase as can be seen from 
\fref{emf_noise_dyn}. The mean noise ranges from 
a few percent to 30\% for the $y$ component of 
the EMF with the dispersion $\sigma$ less than 
30\%. For the $x$ component, the mean residual 
noise has a similar range except around $z\sim 
-0.5$ kpc where it becomes as much as 70\%. The 
dispersion in the noise is also larger at this 
location. At all heights, however the zero value 
is within the 1-$\sigma$ range of the noise 
distribution. These features indicate that while 
the SVD method provides an excellent fit in the 
kinematic phase, it is not providing as good a 
fit in the dynamical phase. At the same time, we 
shall see in Section~\ref{sec:1-d_dynamo} that 
the a 1-D mean field dynamo model using the 
turbulent dynamo coefficients obtained from the
SVD method reproduces reasonably well the 
evolution of the mean magnetic field, not only 
in the kinematic phase but also in the dynamical 
phase.

\section{Comparison of one-D dynamo model with the DNS}
\label{sec:1-d_dynamo}

The validity of the computed profiles of dynamo 
coefficients in the kinematic and dynamical phase 
can also be verified, self-consistently, by 
demonstrating that a 1-D dynamo model using these 
coefficients gives results very similar to the DNS. 
The 1-D dynamo equations are,
\begin{align}
\frac{\partial\overline{{B}}_x}{\partial t}& =
\frac{\partial}{\partial z} \Bigg(
-\left(\overline{U}_z +\alpha_{yx}\right)   \overline{{B}}_x
-\alpha_{yy}    \,                          \overline{{B}}_y
-\eta_{yy}     \,                          \overline{{J}}_y
-\eta_{yx}     \,                          \overline{{J}}_x
\Bigg)\nonumber\\\nonumber
\frac{\partial\overline{{B}}_y}{\partial t} &=
\frac{\partial}{\partial z} \Bigg(
-\left(\overline{U}_z -\alpha_{xy}\right)   \overline{{B}}_y
+\alpha_{xx}      \,                        \overline{{B}}_x
+\eta_{xx}       \,                        \overline{{J}}_x
-\eta_{xy}       \,                        \overline{{J}}_y
\Bigg)\nonumber\\&+q\,\Omega\,\overline{{B}}_x.
\label{e_dynamo}
\end{align}
Note that from $\nabla\cdot{\mathbf{B}}=0$, $
\mean{B}_z=\rm const.$ for the $x-y$ averaged 
mean. We solve \eref{e_dynamo} with a resolution 
of 512 grid points similar to the DNS. Adopting 
a continuous gradient boundary condition for $
\mean{B}_x $ and $\mean{B}_y$, we evolve 
\eref{e_dynamo} on a staggered grid using the 
finite difference method. Initial profiles of 
$\mean{B}_x$ and $\mean{B}_y $ components are 
taken directly from the respective DNS data
averaged over first $150 $\Myr.

The vertical profiles of dynamo coefficients 
used for the first gigayear are as determined 
from the SVD analysis and given in 
\fref{alpha_eta}. For the latter period of 1 to 
$1.5$ \Gyr, we use the linearly interpolated 
profiles of $\alpha_{yy}$ and pumping term 
between black-solid and blue-dashed curves shown 
in \fref{alpha_quench} to roughly mimic the 
transition between kinematic and dynamical 
phases and keep other coefficients the same. For 
the period after $1.5$\Gyr, to simulate the 
dynamical quenching of the coefficients, we 
further replace these with the once shown in
\fref{alpha_quench} with blue-dashed curves and 
keep the rest of the coefficients constant. We 
run this model up to $2.54$\Gyr.

\begin{figure*}
\centering
\includegraphics[width=0.8\linewidth]{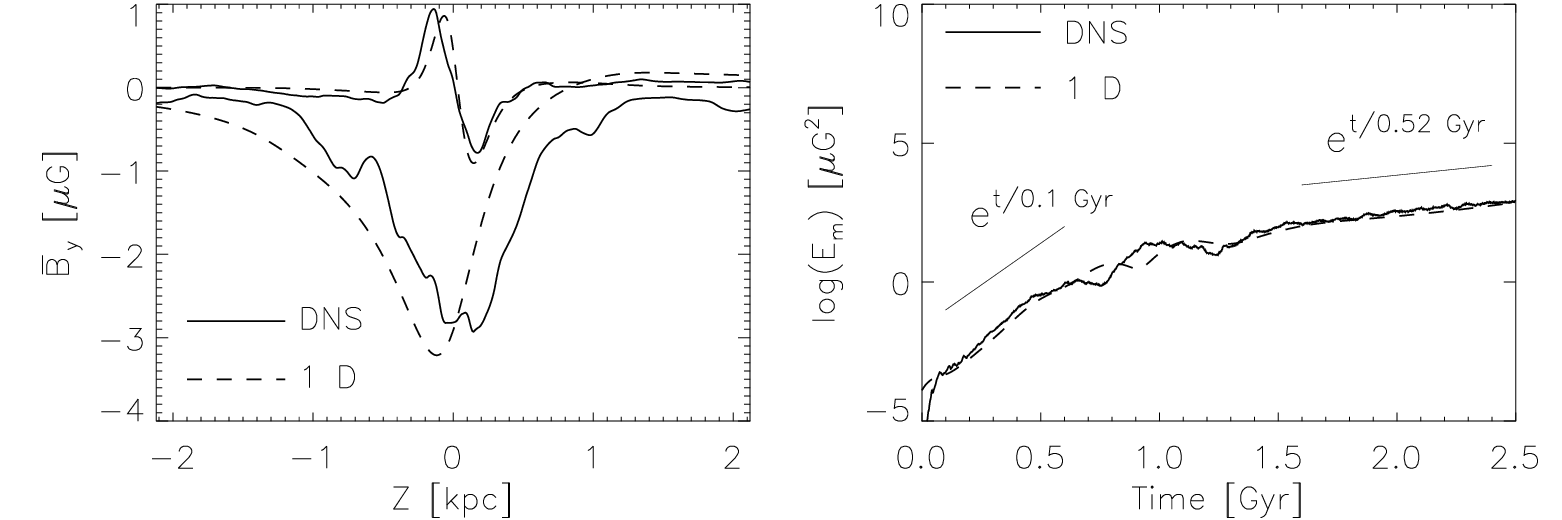}
\caption{\textit{Left Panel:} Black-solid lines 
	show the vertical profile of $\overline{{B
	}}_y$ seen in the DNS, while its dashed 
	counterparts show the same for 1D dynamo 
	model. The approximately symmetric profiles 
	with respect to the midplane corresponds to 
	the $\overline{{B}}_y$ at end of dynamical 
	phase, 2.5 \Gyr, and the antisymmetric ones 
	correspond to 0.8\Gyr. \textit{Right Panel:} 
	The black-solid line represents the time 
	evolution of the mean-magnetic energy 
	expressed in logarithmic units for the DNS,
        while the dashed line represents the same 
	for 1-D model.}
\label{comparison_1}
\end{figure*}

\begin{figure*}
\centering
\includegraphics[width=0.9\linewidth]{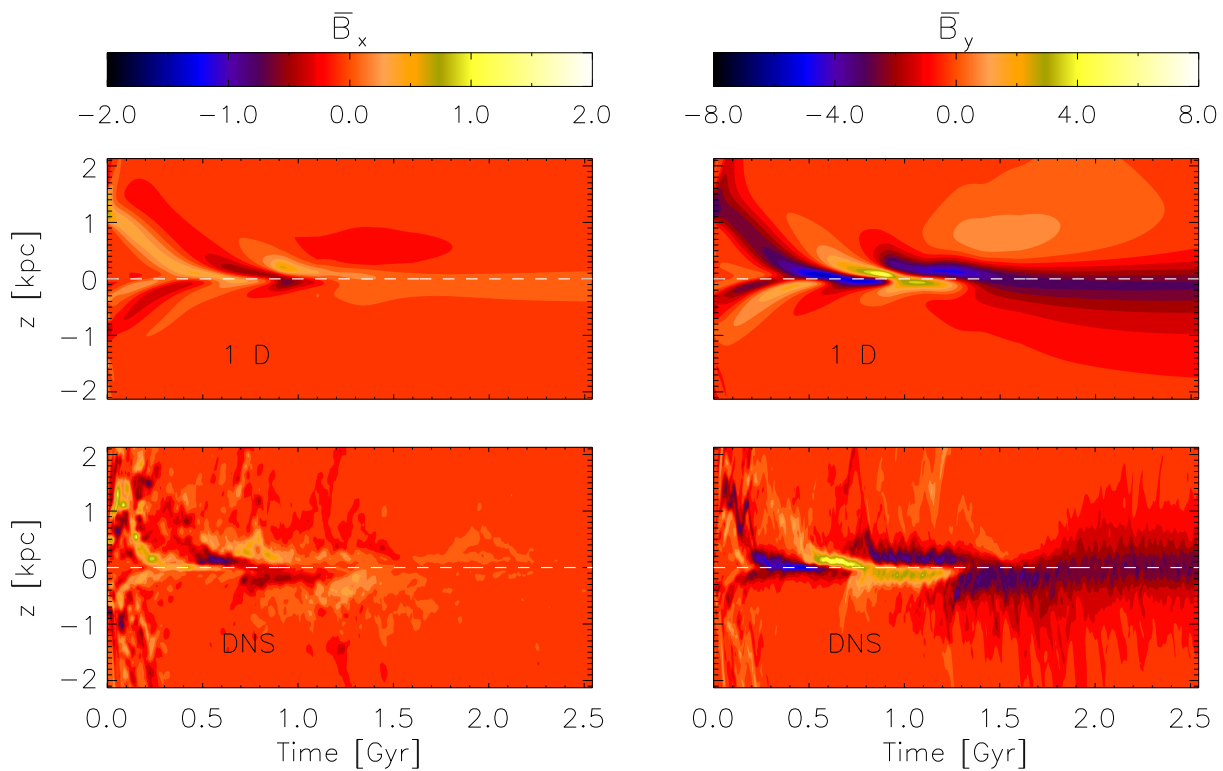}
\caption{\textit{Left Panel:} Shows the time 
	evolution of the vertical profile of 
	$\overline{ B}_x$ seen in the DNS (bottom 
	panel), and in the 1D dynamo model (top 
	panel) \textit{Right Panel:} Shows the same
        but for the $y$ component of mean field. We 
	have normalized the colour code with mean 
	magnetic energy to compensate for the 
	exponential amplification of mean field and 
	make its initial features visible.}
\label{comparison_2}
\end{figure*}

The results of these simulations are shown in 
\fref{comparison_1}, and \fref{comparison_2}. 
After an initial period of $\sim80$\Myr when the 
initial transients decay, the overall evolution 
of magnetic field in 1-D model is reasonably 
consistent with the outcome of the DNS. To 
corroborate this, in \fref{comparison_1} (right 
panel) we first compare the time evolution of 
mean magnetic energies in DNS (shown with solid
line) and 1-D simulations (shown with dashed 
line). We see that mean magnetic energy curves 
from the direct and 1-D simulations, overlap 
closely both in kinematic and dynamical phase. 
This clearly shows the overall similarity in 
the magnetic energy growth, with the e-folding 
time of $\sim100$\Myr in the kinematic and 
$\sim520$\Myr in the dynamical phase as in the 
DNS. Furthermore, in \fref{comparison_2} we 
compare the space-time butterfly diagrams of 
azimuthal field component ($z-t$ evolution). 
This also shows the qualitative similarity with 
which the field profile is reproduced along 
with the reversals and the emergence of final 
symmetric mode. To supplement this, we 
additionally compare in \fref{comparison_1} 
(left panel) the vertical profiles of 
$\overline{B}_y$ from DNS (solid lines) and 1-D 
simulations (dashed lines) at an intermediate 
time of $0.8$\Gyr (when there obtains an 
antisymmetric mode with respect to the 
mid-plane) and near the end of the simulation 
at $2.5$\Gyr, (when a symmetric mode is 
prevalent). This comparison also shows that 
the approximate shape of the $ \overline{B}_y$ 
profile is well replicated in 1-D dynamo 
simulations. Overall, the similarity in the 
evolution of the mean magnetic field in the DNS 
and 1-D models which uses the dynamo 
coefficients determined using SVD method 
supports the robustness of the chosen approach.

\section{Discussion and Conclusions}
\label{discussion}
The determination of turbulent transport 
coefficients in direct simulations which give 
rise to large-scale dynamo action is important 
to understand how the dynamo operates to grow 
and maintain the large-scale magnetic fields. 
Several methods have been suggested in the past, 
including what is known as the test-field 
({TF}) method and the singular value 
decomposition (SVD) method. The SVD is useful 
particularly for post processing analysis of 
simulation data.

We have presented in this paper a SVD analysis of 
the simulation of SNe driven ISM turbulence of 
\citet*{bendre2015dynamo}, which had led to 
large-scale field generation. {TF} results 
for this simulation have been presented there, which
makes it also possible to compare the results 
obtained from these two very different methods.
The profiles of dynamo coefficient tensors
$\alpha_{i\!j}(z) $ and $\eta_{i\!j}(z)$ in the SVD 
method are obtained from the turbulence data
by minimising the least squares of a residual 
vector $R_{i}=\mean{\mathcal{E}}_{i}-\alpha_{i\!j}
\,\mean{B}_{j}+\eta_{i\!j}\,\mean{J}_{j}$.
As a consistency check we verify the efficacy of 
SVD algorithm in \aref{app_mockdata} and by using 
the exact data produced in 1-D simulations, adding 
random white noise up to a level of 50\% of the
actual data, and showing that the SVD effectively 
reconstructs the dynamo coefficient tensors.
The profiles of $\alpha_{i\!j}(z)$ and $\eta_{i\!
j}(z)$ tensors, calculated using the SVD method, 
are shown in \fref{alpha_eta} for the kinematic 
phase and \fref{alpha_quench_all} for the 
dynamical phase when the dynamo growth has 
decreased. We also show that the turbulent EMF 
components predicted using the reconstructed $
\alpha_{i\!j}$ and $\eta_{i\!j}$ tensors match quite 
well with actual DNS data for the kinematic phase,
with very little residual noise. The match is not 
as good for the dynamical phase. However, we show 
that the evolution of the mean magnetic fields, 
predicted by solving the 1-D mean-field dynamo
equations using the reconstructed dynamo 
coefficients match remarkably well, with that 
determined from the DNS, as can be seen from 
\fref{comparison_1} and \fref{comparison_2}. Thus 
it seems that the SVD does indeed give a reasonable
reconstruction of the dynamo coefficients.

The predicted magnitude of $\alpha_{yy}$, crucial 
for regenerating $\mean{B}_x$ from $\mean{B}_y$ in 
the $\alpha-\Omega$ dynamo, is zero at the midplane 
(as expected) and rises with $z$ to attain a 
maximum of about $3 {\rm\km\s^{-1}}$ by $z= 1$ \kpc.
We also predict a turbulent pumping term $\gamma
\sim10$\kms at the height of a \kpc, that acts to 
transport mean fields towards the equator, against 
the outward advection by the vertical velocity 
$\overline{U}_z$. The diagonal components of 
turbulent diffusion tensor, as recovered by the 
SVD method, are much more noisy, approximately 
inverted bell shaped, with a maximum turbulent 
diffusivity of $\simeq 10^{26} {\rm \cm^2 \s^{-1}}$
at a distance of a \kpc from the disk midplane. 
These numbers compare favorably with that expected 
on basis of simple estimates for galactic dynamos
\citep{anvar_2004}. We find that as the mean field 
becomes stronger, both $\alpha_{yy}$ and $\gamma$ 
get suppressed, but other coefficients remain 
largely unaltered.

The vertical profiles of all dynamo coefficients 
constructed in SVD analysis (\fref{alpha_eta}) are 
qualitatively similar to their {TF} counterparts 
(shown in \fref{alpha_eta_tf}) during the kinematic 
phase. {In this phase, they are also similar to 
that obtained from a simple regression method of 
{BS02}, as shown in 
\fref{fig:alpha_eta_regression}.} The amplitude of 
$\alpha_{xx}$ however is larger by a factor of 3 as 
determined in the SVD analysis {compared to the 
{TF} method}, while $\eta_{xx}$ and $\eta_{yy}$ 
are substantially smaller. We stress that the 
SVD method is likely most sensitive to vertical 
gradients in the mean fields on relatively {smaller} 
scales. This is because it is restricted to the length 
scales that are actually sampled from the fields 
present in the DNS, which are confined to a few 
hundred parsec around the midplane. In contrast, the 
{TF} method (in its form used here) is 
essentially a spectral method, where one is free to 
choose the length scale probed. It has to be noted 
that the particular choice was to have test fields 
that vary on the largest vertical scales accessible 
in the tall simulation domain. \citet{GP15} have 
demonstrated the scale-dependence of the {TF} 
coefficients for the case of magnetorotational 
turbulence, where coefficients were found to decay 
from their peak value at the largest available scales 
by a factor of a few, when approaching the smallest 
scales accessible to the simulation. This 
scale-dependence may in fact fully explain the 
tension between the SVD and {TF} results, which 
can be tested by running the latter method at higher 
vertical wavenumber.

In the dynamically quenched phase, the {TF} 
method predicts not only suppression of $\alpha_{yy}$ 
and $\gamma$ but also the turbulent diffusion tensor, 
with the latter feature not having been {clearly} 
obtained in the SVD analysis. Nevertheless, the 1-D 
model using the {TF} results had also matched the 
evolution of the mean magnetic fields obtained in the 
DNS \citep*{bendre2015dynamo}. This seems to indicate 
a degeneracy in the determination of the magnitudes of 
turbulent transport coefficients using different methods.
A potential explanation for this feature is the existence 
of partial correlations between some of the parameters, 
as explicitly shown in \fref{covar}.

Our work here has demonstrated that the SVD method is a 
self-consistent and useful way of determining turbulent
transport coefficients. In contrast to the {TF}
method, it is computationally less expensive since
it is used merely as a post-processing tool. However, as 
a potential downside of the SVD method, some of the
determined parameters can also get correlated if the 
different components of the mean fields and currents 
have definite correlations. The reconstruction of 
coefficients which couple to the current is also more 
noisy in the SVD method. These latter issues are tacitly 
avoided in {TF} method as they study the inductive
response to known functional forms of additional mean 
test fields. The {TF} method itself is believed to 
have difficulties in dealing with a strong small-scale 
dynamo, where small scale fields are generated independent 
of the mean magnetic field. Here such fields will merely 
appear as an extra noise term to be recovered 
self-consistently in the SVD reconstruction. It will be 
important to study a case where both large and 
small-scale dynamos are active (e.g., the case in 
\citet{BSB19}), with the SVD method. Also of interest 
will be to recover the dynamo coefficients when the mean 
field is defined differently, like in the filtering
approach (eg. \cite{gent_2013}). In addition, it would 
be useful to be able to implement Bayesian priors 
for the dynamo coefficients, perhaps using the 
information field theory approach 
\citep[e.g.,][]{IFT_torsten}, while doing the 
least-square minimisation of the data; a study which is 
left for the future.

\section*{Acknowledgements}

We used the NIRVANA code version 3.3, developed by 
Udo Ziegler at the Leibniz-Institut f\"ur Astrophysik 
Potsdam (AIP). For computations we used Leibniz Computer 
Cluster, also at AIP. We thank Dipankar Bhattacharya, 
Torsten En{\ss}lin and Anvar Shukurov for very 
insightful discussions.



\bibliographystyle{mnras}
\bibliography{references} 

\begin{thebibliography}{}
\makeatletter
\relax
\def\mn@urlcharsother{\let\do\@makeother \do\$\do\&\do\#\do\^\do\_\do\%\do\~}
\def\mn@doi{\begingroup\mn@urlcharsother \@ifnextchar [ {\mn@doi@}
  {\mn@doi@[]}}
\def\mn@doi@[#1]#2{\def\@tempa{#1}\ifx\@tempa\@empty \href
  {http://dx.doi.org/#2} {doi:#2}\else \href {http://dx.doi.org/#2} {#1}\fi
  \endgroup}
\def\mn@eprint#1#2{\mn@eprint@#1:#2::\@nil}
\def\mn@eprint@arXiv#1{\href {http://arxiv.org/abs/#1} {{\tt arXiv:#1}}}
\def\mn@eprint@dblp#1{\href {http://dblp.uni-trier.de/rec/bibtex/#1.xml}
  {dblp:#1}}
\def\mn@eprint@#1:#2:#3:#4\@nil{\def\@tempa {#1}\def\@tempb {#2}\def\@tempc
  {#3}\ifx \@tempc \@empty \let \@tempc \@tempb \let \@tempb \@tempa \fi \ifx
  \@tempb \@empty \def\@tempb {arXiv}\fi \@ifundefined
  {mn@eprint@\@tempb}{\@tempb:\@tempc}{\expandafter \expandafter \csname
  mn@eprint@\@tempb\endcsname \expandafter{\@tempc}}}

\bibitem[\protect\citeauthoryear{Beck}{Beck}{2012}]{Beck2012}
Beck R.,  2012, \mn@doi [Space Science Reviews] {10.1007/s11214-011-9782-z},
  166, 215

\bibitem[\protect\citeauthoryear{{Beck} \& {Wielebinski}}{{Beck} \&
  {Wielebinski}}{2013}]{beck_wielebinski}
{Beck} R.,  {Wielebinski} R.,  2013, {Magnetic Fields in Galaxies}.
Springer Berlin Heidelberg, p.~641, \mn@doi{10.1007/978-94-007-5612-0_13}

\bibitem[\protect\citeauthoryear{Beck, Brandenburg, Moss, Shukurov  \&
  Sokoloff}{Beck et~al.}{1996}]{beck_1996}
Beck R.,  Brandenburg A.,  Moss D.,  Shukurov A.,   Sokoloff D.,  1996, \mn@doi
  [Annual Review of Astronomy and Astrophysics]
  {10.1146/annurev.astro.34.1.155}, 34, 155

\bibitem[\protect\citeauthoryear{Bendre}{Bendre}{2016}]{Bendre2016}
Bendre A.~B.,  2016, doctoralthesis, Universit{\"a}t Potsdam

\bibitem[\protect\citeauthoryear{Bendre, Gressel  \& Elstner}{Bendre
  et~al.}{2015}]{bendre2015dynamo}
Bendre A.,  Gressel O.,   Elstner D.,  2015, Astronomische Nachrichten, 336,
  991

\bibitem[\protect\citeauthoryear{{Bhat}, {Subramanian}  \&
  {Brandenburg}}{{Bhat} et~al.}{2019}]{BSB19}
{Bhat} P.,  {Subramanian} K.,   {Brandenburg} A.,  2019, arXiv e-prints, \href
  {https://ui.adsabs.harvard.edu/abs/2019arXiv190508278B} {p. arXiv:1905.08278}

\bibitem[\protect\citeauthoryear{{Brandenburg}}{{Brandenburg}}{2005}]{Bran05}
{Brandenburg} A.,  2005, \mn@doi [Astronomische Nachrichten]
  {10.1002/asna.200510414}, \href
  {http://adsabs.harvard.edu/abs/2005AN....326..787B} {326, 787}

\bibitem[\protect\citeauthoryear{Brandenburg}{Brandenburg}{2009}]{Brandenburg2009}
Brandenburg A.,  2009, \mn@doi [Space Science Reviews]
  {10.1007/s11214-009-9490-0}, 144, 87

\bibitem[\protect\citeauthoryear{Brandenburg}{Brandenburg}{2018}]{brandenburg_2018}
Brandenburg A.,  2018, \mn@doi [Journal of Plasma Physics]
  {10.1017/S0022377818000806}, 84, 735840404

\bibitem[\protect\citeauthoryear{{Brandenburg} \& {Sokoloff}}{{Brandenburg} \&
  {Sokoloff}}{2002}]{BranSok02}
{Brandenburg} A.,  {Sokoloff} D.,  2002, \mn@doi [Geophysical \& Astrophysical
  Fluid Dynamics] {10.1080/03091920290032974}, \href
  {http://adsabs.harvard.edu/abs/2002GApFD..96..319B} {96, 319}

\bibitem[\protect\citeauthoryear{{Brandenburg} \& {Subramanian}}{{Brandenburg}
  \& {Subramanian}}{2005}]{BS05}
{Brandenburg} A.,  {Subramanian} K.,  2005, \mn@doi [Physics Reports]
  {10.1016/j.physrep.2005.06.005}, \href
  {https://ui.adsabs.harvard.edu/abs/2005PhR...417....1B} {417, 1}

\bibitem[\protect\citeauthoryear{{Cattaneo} \& {Hughes}}{{Cattaneo} \&
  {Hughes}}{1996}]{CH96}
{Cattaneo} F.,  {Hughes} D.~W.,  1996, \mn@doi [Physical Review E]
  {10.1103/PhysRevE.54.R4532}, \href
  {http://adsabs.harvard.edu/abs/1996PhRvE..54.4532C} {54, 4532}

\bibitem[\protect\citeauthoryear{{En{\ss}lin}}{{En{\ss}lin}}{2019}]{IFT_torsten}
{En{\ss}lin} T.~A.,  2019, \mn@doi [Annalen der Physik]
  {10.1002/andp.201800127}, \href
  {https://ui.adsabs.harvard.edu/abs/2019AnP...53100127E} {531, 1800127}

\bibitem[\protect\citeauthoryear{{Fletcher}}{{Fletcher}}{2010}]{fletcher_nearby_2010}
{Fletcher} A.,  2010, in {Kothes} R.,  {Landecker} T.~L.,   {Willis} A.~G.,
  eds,  Astronomical Society of the Pacific Conference Series Vol. 438, The
  Dynamic Interstellar Medium: A Celebration of the Canadian Galactic Plane
  Survey. p.~197 (\mn@eprint {arXiv} {1104.2427})

\bibitem[\protect\citeauthoryear{{Gent}, {Shukurov}, {Sarson}, {Fletcher}  \&
  {Mantere}}{{Gent} et~al.}{2013}]{gent_2013}
{Gent} F.~A.,  {Shukurov} A.,  {Sarson} G.~R.,  {Fletcher} A.,   {Mantere}
  M.~J.,  2013, \mn@doi [\mnras] {10.1093/mnrasl/sls042}, \href
  {https://ui.adsabs.harvard.edu/abs/2013MNRAS.430L..40G} {430, L40}

\bibitem[\protect\citeauthoryear{{Gressel} \& {Pessah}}{{Gressel} \&
  {Pessah}}{2015}]{GP15}
{Gressel} O.,  {Pessah} M.~E.,  2015, \mn@doi [\apj]
  {10.1088/0004-637X/810/1/59}, \href
  {https://ui.adsabs.harvard.edu/abs/2015ApJ...810...59G} {810, 59}

\bibitem[\protect\citeauthoryear{{Gressel}, {Elstner}, {Ziegler}  \&
  {R{\"u}diger}}{{Gressel} et~al.}{2008}]{gressel_2008}
{Gressel} O.,  {Elstner} D.,  {Ziegler} U.,   {R{\"u}diger} G.,  2008, \mn@doi
  [Astronomy and Astrophysics] {10.1051/0004-6361:200810195}, \href
  {https://ui.adsabs.harvard.edu/abs/2008A&A...486L..35G} {486, L35}

\bibitem[\protect\citeauthoryear{{Gressel}, {Bendre}  \& {Elstner}}{{Gressel}
  et~al.}{2013}]{gressel2012magnetic}
{Gressel} O.,  {Bendre} A.,   {Elstner} D.,  2013, \mn@doi [\mnras]
  {10.1093/mnras/sts356}, \href
  {http://adsabs.harvard.edu/abs/2013MNRAS.429..967G} {429, 967}

\bibitem[\protect\citeauthoryear{{K\"apyl\"a}, {Korpi}  \&
  {Brandenburg}}{{K\"apyl\"a} et~al.}{2009}]{kapala_test}
{K\"apyl\"a} P.~J.,  {Korpi} M.~J.,   {Brandenburg} A.,  2009, \mn@doi
  [Astronomy and Astrophysics] {10.1051/0004-6361/200811498}, 500, 633

\bibitem[\protect\citeauthoryear{{Kowal}, {Otmianowska-Mazur}  \&
  {Hanasz}}{{Kowal} et~al.}{2006}]{Kowal06}
{Kowal} G.,  {Otmianowska-Mazur} K.,   {Hanasz} M.,  2006, \mn@doi [\aap]
  {10.1051/0004-6361:20053582}, \href
  {http://adsabs.harvard.edu/abs/2006A%26A...445..915K} {445, 915}

\bibitem[\protect\citeauthoryear{Krause et~al.,}{Krause
  et~al.}{2018}]{krause2018chang}
Krause M.,  et~al., 2018, Astronomy \& Astrophysics, 611, A72

\bibitem[\protect\citeauthoryear{{Kuijken} \& {Gilmore}}{{Kuijken} \&
  {Gilmore}}{1989a}]{gravity_1}
{Kuijken} K.,  {Gilmore} G.,  1989a, \mn@doi [\mnras]
  {10.1093/mnras/239.2.571}, \href
  {http://adsabs.harvard.edu/abs/1989MNRAS.239..571K} {239, 571}

\bibitem[\protect\citeauthoryear{{Kuijken} \& {Gilmore}}{{Kuijken} \&
  {Gilmore}}{1989b}]{gravity_2}
{Kuijken} K.,  {Gilmore} G.,  1989b, \mn@doi [\mnras]
  {10.1093/mnras/239.2.605}, \href
  {http://adsabs.harvard.edu/abs/1989MNRAS.239..605K} {239, 605}

\bibitem[\protect\citeauthoryear{{Kuijken} \& {Gilmore}}{{Kuijken} \&
  {Gilmore}}{1989c}]{gravity_3}
{Kuijken} K.,  {Gilmore} G.,  1989c, \mn@doi [\mnras]
  {10.1093/mnras/239.2.651}, \href
  {http://adsabs.harvard.edu/abs/1989MNRAS.239..651K} {239, 651}

\bibitem[\protect\citeauthoryear{Mandel}{Mandel}{1982}]{mendel_svd}
Mandel J.,  1982, \mn@doi [The American Statistician]
  {10.1080/00031305.1982.10482771}, 36, 15

\bibitem[\protect\citeauthoryear{{Moffatt}}{{Moffatt}}{1978}]{Mof78}
{Moffatt} H.~K.,  1978, {Magnetic Field Generation in Electrically Conducting
  Fluids}.
Cambridge University Press, Cambridge

\bibitem[\protect\citeauthoryear{Press, Teukolsky, Vetterling  \&
  Flannery}{Press et~al.}{1992}]{recepies}
Press W.~H.,  Teukolsky S.~A.,  Vetterling W.~T.,   Flannery B.~P.,  1992,
  Numerical Recipes in C (2Nd Ed.): The Art of Scientific Computing.
Cambridge University Press, New York, NY, USA

\bibitem[\protect\citeauthoryear{Racine, Charbonneau, Ghizaru, Bouchat  \&
  Smolarkiewicz}{Racine et~al.}{2011}]{racine2011mode}
Racine {\'E}.,  Charbonneau P.,  Ghizaru M.,  Bouchat A.,   Smolarkiewicz
  P.~K.,  2011, The Astrophysical Journal, 735, 46

\bibitem[\protect\citeauthoryear{{R{\"a}dler}}{{R{\"a}dler}}{1969}]{Radler69}
{R{\"a}dler} K.-H.,  1969, Veroeffentlichungen der Geod.~Geophys, \href
  {http://adsabs.harvard.edu/abs/1969VeGG...13..131R} {13, 131}

\bibitem[\protect\citeauthoryear{{R{\"a}dler}}{{R{\"a}dler}}{2014}]{radler2014}
{R{\"a}dler} K.~H.,  2014, arXiv e-prints, \href
  {https://ui.adsabs.harvard.edu/abs/2014arXiv1402.6557R} {p. arXiv:1402.6557}

\bibitem[\protect\citeauthoryear{{S{\'a}nchez-Salcedo}, {V{\'a}zquez-Semadeni}
  \& {Gazol}}{{S{\'a}nchez-Salcedo} et~al.}{2002}]{radiative_cooling}
{S{\'a}nchez-Salcedo} F.~J.,  {V{\'a}zquez-Semadeni} E.,   {Gazol} A.,  2002,
  \mn@doi [\apj] {10.1086/342223}, \href
  {http://adsabs.harvard.edu/abs/2002ApJ...577..768S} {577, 768}

\bibitem[\protect\citeauthoryear{{Schrinner}, {R{\"a}dler}, {Schmitt},
  {Rheinhardt}  \& {Christensen}}{{Schrinner} et~al.}{2005}]{schriner_test}
{Schrinner} M.,  {R{\"a}dler} K.-H.,  {Schmitt} D.,  {Rheinhardt} M.,
  {Christensen} U.,  2005, \mn@doi [Astronomische Nachrichten]
  {10.1002/asna.200410384}, \href
  {http://adsabs.harvard.edu/abs/2005AN....326..245S} {326, 245}

\bibitem[\protect\citeauthoryear{{Schrinner}, {R{\"a}dler}, {Schmitt},
  {Rheinhardt}  \& {Christensen}}{{Schrinner} et~al.}{2007}]{schriner_test1}
{Schrinner} M.,  {R{\"a}dler} K.-H.,  {Schmitt} D.,  {Rheinhardt} M.,
  {Christensen} U.~R.,  2007, \mn@doi [Geophysical and Astrophysical Fluid
  Dynamics] {10.1080/03091920701345707}, \href
  {http://adsabs.harvard.edu/abs/2007GApFD.101...81S} {101, 81}

\bibitem[\protect\citeauthoryear{{Shukurov}}{{Shukurov}}{2004}]{anvar_2004}
{Shukurov} A.,  2004, arXiv e-prints, \href
  {https://ui.adsabs.harvard.edu/abs/2004astro.ph.11739S} {pp
  astro--ph/0411739}

\bibitem[\protect\citeauthoryear{Shukurov}{Shukurov}{2005}]{shukurov_2005}
Shukurov A.,  2005, Mesoscale Magnetic Structures in Spiral Galaxies.
Springer Berlin Heidelberg, Berlin, Heidelberg, pp 113--135,
  \mn@doi{10.1007/3540313966_6}, \url {https://doi.org/10.1007/3540313966_6}

\bibitem[\protect\citeauthoryear{Simard, Charbonneau  \& Dub{\'e}}{Simard
  et~al.}{2016}]{simard2016characterisation}
Simard C.,  Charbonneau P.,   Dub{\'e} C.,  2016, Advances in Space Research,
  58, 1522

\bibitem[\protect\citeauthoryear{Sur, Subramanian  \& Brandenburg}{Sur
  et~al.}{2007}]{sur2007kinetic}
Sur S.,  Subramanian K.,   Brandenburg A.,  2007, Monthly Notices of the Royal
  Astronomical Society, 376, 1238

\bibitem[\protect\citeauthoryear{{Tobias} \& {Cattaneo}}{{Tobias} \&
  {Cattaneo}}{2013}]{angstrom}
{Tobias} S.~M.,  {Cattaneo} F.,  2013, \mn@doi [Journal of Fluid Mechanics]
  {10.1017/jfm.2012.575}, \href
  {https://ui.adsabs.harvard.edu/abs/2013JFM...717..347T} {717, 347}

\bibitem[\protect\citeauthoryear{{Warnecke}, {Rheinhardt}, {Tuomisto},
  {K{\"a}pyl{\"a}}, {K{\"a}pyl{\"a}}  \& {Brandenburg}}{{Warnecke}
  et~al.}{2016}]{War17}
{Warnecke} J.,  {Rheinhardt} M.,  {Tuomisto} S.,  {K{\"a}pyl{\"a}} P.~J.,
  {K{\"a}pyl{\"a}} M.~J.,   {Brandenburg} A.,  2016, preprint, \href
  {http://adsabs.harvard.edu/abs/2016arXiv160103730W} {} (\mn@eprint {arXiv}
  {1601.03730})

\bibitem[\protect\citeauthoryear{{Ziegler}}{{Ziegler}}{2008}]{nirvana}
{Ziegler} U.,  2008, \mn@doi [Computer Physics Communications]
  {10.1016/j.cpc.2008.02.017}, \href
  {http://adsabs.harvard.edu/abs/2008CoPhC.179..227Z} {179, 227}

\makeatother
\end{thebibliography}



\appendix

\section{Testing the SVD method using mock data}
\label{app_mockdata}

\begin{figure}
\centering
\includegraphics[width=\linewidth]{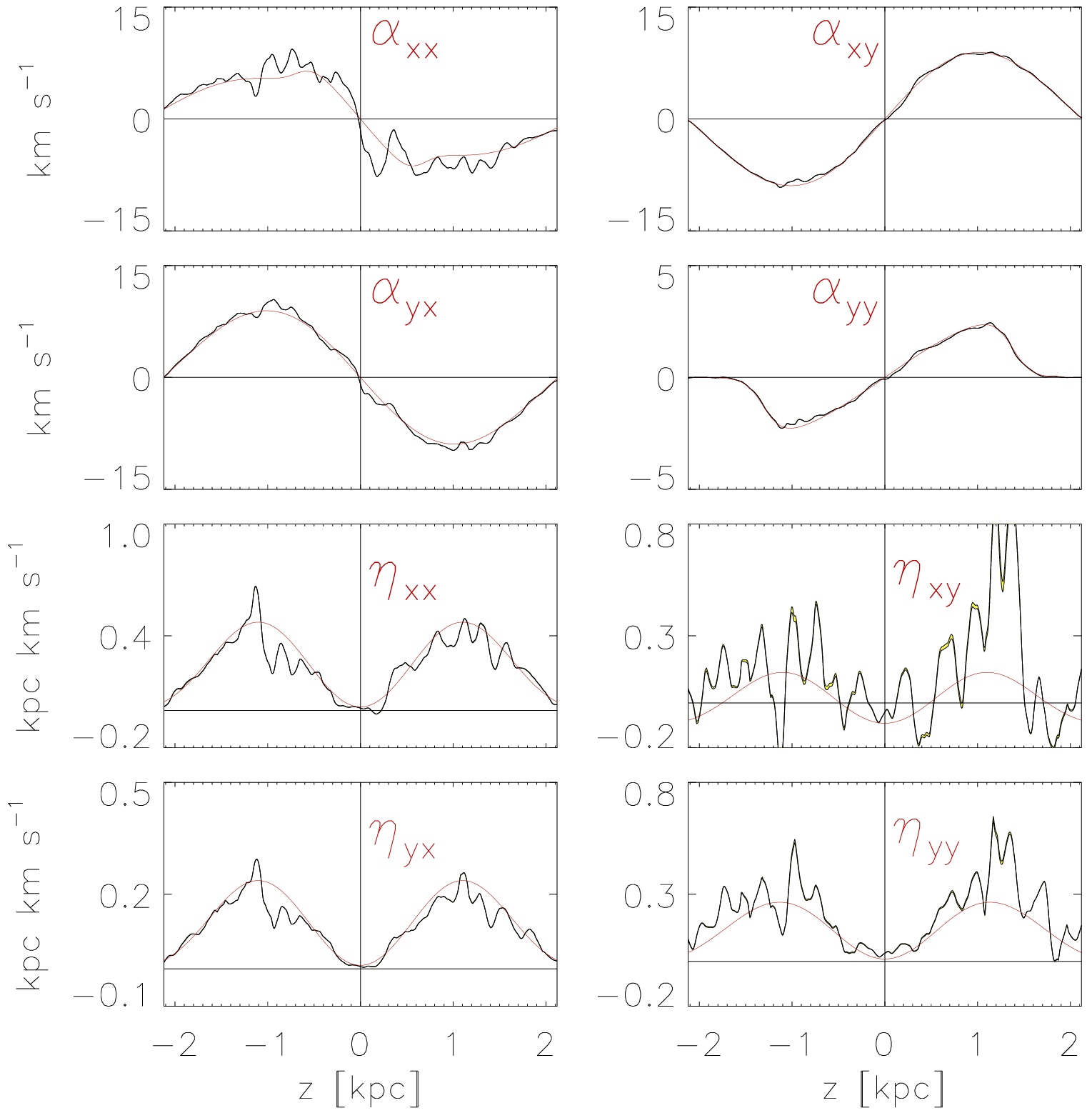}
\caption{Black-lines show the vertical profiles of all
	dynamo coefficients reconstructed from 1D dynamo 
	data with added noise level of 50\% to the actual 
	data. Red-lines show the dynamo coefficients actually 
	used in the 1D Dynamo simulations.}
\label{noise_50_diff}
\end{figure}

To verify the robustness and predictive ability of the 
SVD method for the problem of mean-field dynamo, we use 
mock data of the mean magnetic fields produced in 1D 
dynamo simulations using the profiles of dynamo 
coefficients (described in \sref{sec:1-d_dynamo}).
We extract the 
time series data of $ \overline{B}_x \left(z,t\right)$ 
and $\overline{B}_y \left( z,t \right) $ and add the 
various levels of normally distributed random noise over 
these time series. We denote these noisy components 
$ \overline{B}^n_x$ and $\overline{B}^n_y $. To then 
compute the time series of noisy mean current components 
$  \overline{ J }_x^n$ and $ \overline{J}_y^n$ we filter 
the data $\overline{B}^n_x\left(z,t\right)$ and $ 
\overline{B }^n_y \left( z,t \right)$ using moving
windows of width $ \sim80$\pc in $z$ direction and $\sim 
60$ \Myr in $ t  $ direction, and calculate the components 
of$ \nabla\times \overline{ \mathbf{B}}^n$. Similarly we 
also construct the time series of $\overline{\mathcal{E}}_x$ 
and $\overline{\mathcal{ E }}_y $ from the exact profiles of 
$\overline{ B}_x $, $\overline{B }_y$, $\overline{ J }_x $,
$ \overline{   J }_y $ and the dynamo coefficients obtained
by SVD. By adding the same levels of noise over $\overline{
\mathcal{E}}$, and then filtering with same window size the
time series $ \overline{ \mathcal{  E }}^n $ components are
constructed. Using these noisy time series $     \overline{
\mathbf{B}}^n$, $\overline{ \mathbf{J}}^n$ and $ \overline{
\mathcal{E}}^n$, we subsequently follow the steps described
in \sref{sec:svd_method} to pseudo-invert the design matrix
$\mathbf{ A } $ at each $z$ and reconstruct the profiles of
all dynamo coefficients, along with their variances. In
\fref{noise_50_diff} we show the reconstructed profiles of
the dynamo coefficients with a level of added noise 
equaling to approximately $50\%$ of the actual data (with 
black lines). {Here we point out that noise dispersion 
in the actual DNS data in its kinematic phase is almost an 
order of magnitude smaller than the value we use here. It 
is however almost similar to the expected noise in 
dynamical phase (see \fref{emf_noise_kin} and 
\fref{emf_noise_dyn}).} The red curves {in 
\fref{noise_50_diff}} show actual profiles of dynamo 
coefficients for non-noisy data set. 

{It can be seen from the figure that the predictions of
$\alpha$ tensor components agree remarkably well with the 
input profiles. They are considerably better than the
predictions of $\eta_{ij}$, especially for the off diagonal
component $\eta_{xy}$. We also find that if we neglect the 
off diagonal
components of $\eta$ tensor altogether, this further improves 
the SVD predictions for both $\alpha$ and diagonal elements of
$\eta$}. Qualitative consistency of the reconstructed profiles
with the actual profiles justifies the robustness of SVD
algorithm.

\begin{figure}
\centering
\includegraphics[width=\linewidth]{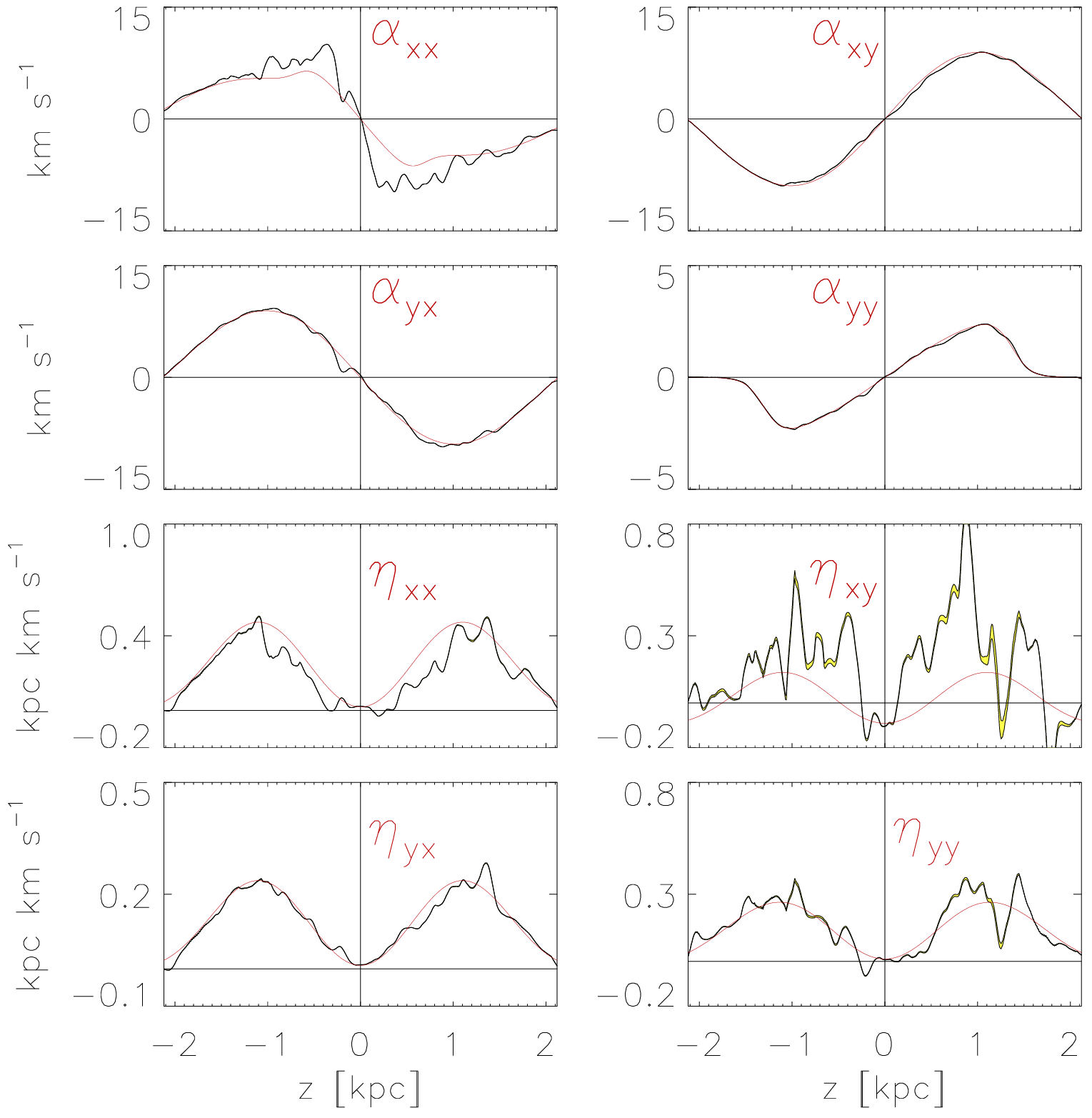}
\caption{Same as \fref{noise_50_diff}, but with currents
components calculated using \eref{fourier_current}}
\label{noise_50_fou}
\end{figure}

One of the main sources of errors in determining the
dynamo coefficients using SVD algorithm is the enhancement
of noise when determining $\overline{\mathbf{J}}^n$ due to
taking the derivative of the noisy mean field data. Large 
relative errors are introduced in the derivatives of 
$\overline{\mathbf{  B } } $ where it is expected to have 
a negligible gradient. To circumvent this issue, we also 
define the components of $ \mean{\mathbf{J}}^n$ as the 
discrete inverse Fourier transforms of $\mean{J}_x^n\left(
k\right)$ and $\mean{J}_y^n\left(k\right)$ defined by
\begin{align}
    \mean{J}_x^n \left(k\right) &= \frac{2 \pi i k}{N}\sum_{z} \mean{B}_y^n\left(z\right) \, \mathrm{exp}\left(\frac{-2 \pi i k z}{N} \right) \nonumber \\
    \mean{J}_y^n \left(k\right) &= - \frac{2 \pi i k}{N}\sum_{z} \mean{B}_x^n\left(z\right) \, \mathrm{exp}\left(\frac{-2 \pi i k z}{N} \right).
    \label{fourier_current}
\end{align}
Resulting time series of $\mean{\mathbf{J}}^n$, $\mean{\mathbf{B}}^n$
and $\mean{\emf}^n$ is then used to determine the dynamo
coefficients corresponding to noisy data, using the SVD
method described above. This exercise also yields the
results consistent with the actual values of dynamo
coefficients even with added noise level of $\sim
50\%$ of actual data, as shown in \fref{noise_50_fou}.
These tests with mock data show that the dynamo coefficients
can be reasonably well recovered using the SVD method.
It also shows that the recovering the $\eta_{i\!j}$ tensor
accurately is more difficult compared to recovery of the
$\alpha_{i\!j}$ tensor.

\section{Test-Field Results}
\label{alphas_tf}

\begin{figure}
\centering
\includegraphics[width=\linewidth]{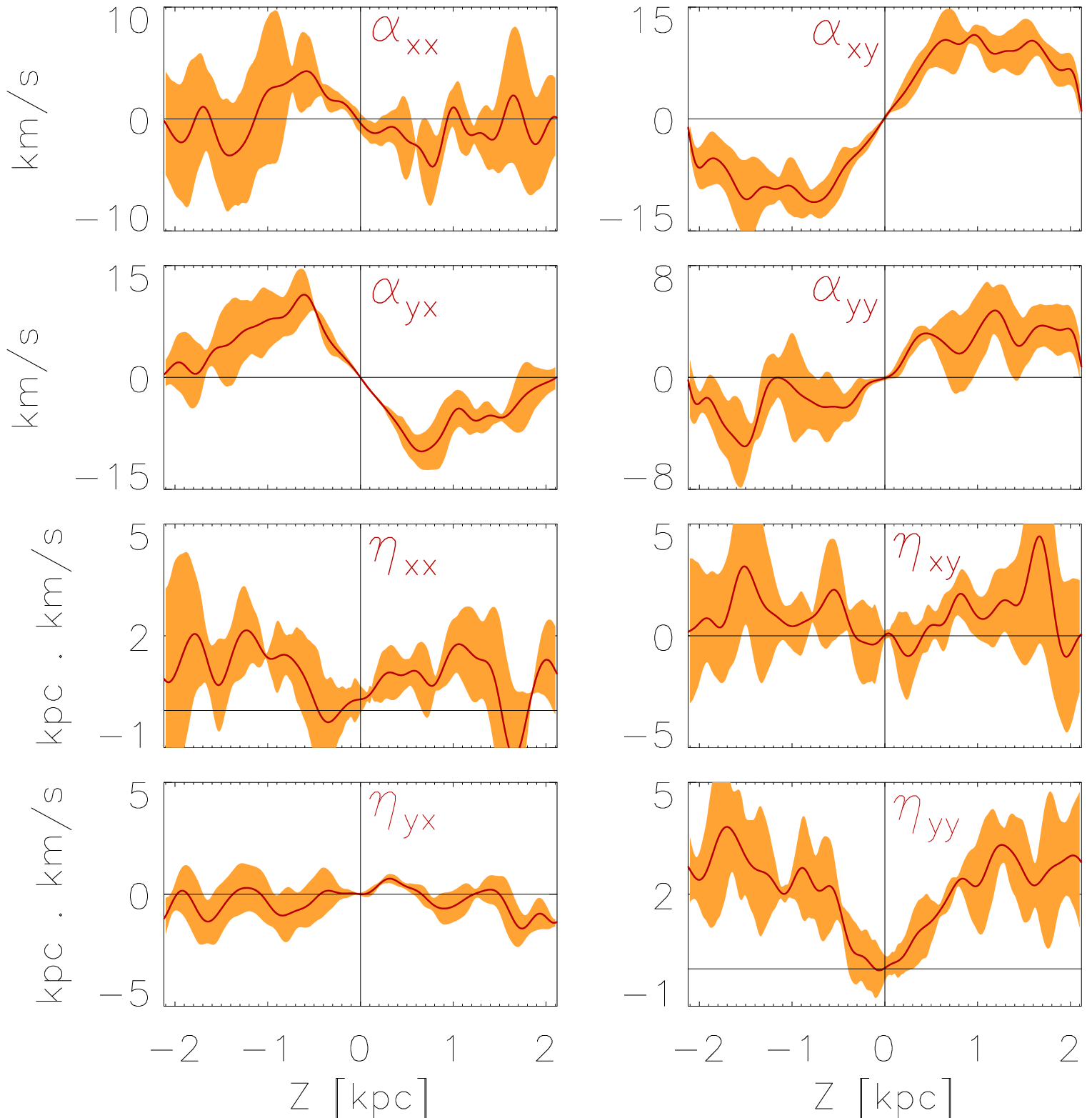}
\caption{Plotted here are the vertical profiles of all 
	dynamo coefficients obtained from {TF} in 
	red-solid lines and corresponding 1-$\sigma$ error 
	estimates shown by the overlayed orange shaded 
	areas. These results correspond to the exponential 
	amplification phase of mean fields $t \in (0.1:0.8
	)$ \Gyr, similar to the SVD results presented in 
	\fref{alpha_eta}. Vertical profiles are also 
	Fourier filtered such that all the fluctuations 
	below the estimated turbulent correlation length 
	scales of $\sim100$ \pc are eliminated.}
\label{alpha_eta_tf}
\end{figure}

In order to solve the system represented by \eref{e1}
for $\alpha$ and $\eta$ tensors, in our previous work
we use the {TF} method discussed by
\citet{brandenburg_2018}. A general idea of the {TF} 
is as follows: Since the \eref{e1} is an under-determined
system with two equations and eight unknowns, one
needs sufficient number of independent equations to
invert it. In the {TF} method, this is achieved 
by solving the induction equation for additional
passive test fields with a well defined functional
form along with the DNS, as described briefly in 
Section~\ref{intro}. Fluctuations in these fields
are then computed as functions of space and time
and used to compute the {TF} EMF components.
With this additional data \eref{e1} is inverted to
compute the unknown $ \alpha $ and $ \eta$ tensors.
Analysis of the DNS model used in this paper (model
Q) based on the {TF} results is discussed in
details in \citet*{bendre2015dynamo}. For the sake of
comparison with the SVD analysis, in \fref{alpha_eta_tf} 
we plot the vertical profiles all dynamo coefficients 
obtained from the {TF} analysis of the same
model in solid red lines, along with 1-$\sigma$ error
estimates represented by the orange shaded regions.

Here we have presented the {TF} results only in the
kinematic regime of field evolution. Changes in these
profiles in the presence of dynamically significant mean
fields are discussed in \citet*{bendre2015dynamo} and in
Section 5.3.2 of \citet{Bendre2016}. Qualitative trends
of these profiles of dynamo coefficients match well with
SVD outcomes as shown in \fref{alpha_eta}. Magnitudes
of $\alpha_{xx}$ and all $\eta_{i\!j}$ coefficients, 
however are predicted to be smaller in the SVD analysis.
Nevertheless the magnitudes $\alpha_{yy}$ important
for the the generation of mean poloidal field from the
toroidal one, and ($\alpha_{xy}$, $\alpha_{yx}$) which
determine the $\gamma$-effect, predicted by both methods
match reasonably well, within the 1-$\sigma$ intervals.

\section{Results from a simple regression method}
\label{sec:regr}

We briefly present results from a statistical method, 
first introduced by {BS02}, that notably uses the 
same data basis as the SVD method explored here. The 
method acknowledges that the inversion of the EMF 
parametrisation is under-constrained and proceeds by 
building statistical moments with the right-hand-side 
variable, i.e., the magnetic field and current.

For direct comparison with our SVD result, we have 
implemented the local (i.e., non-scale dependent) 
formulation without the assumption of a diagonal 
diffusion tensor, as described in detail in section 
4.1 of {BS02}. We only briefly summarise the 
approach here. The x and y portions of the matrix 
equation to be inverted read
\begin{equation}
  \mathbf{E}^{(i)}(z) = \mathcal{\underline{M}}(z)\;\mathcal{C}^{(i)}(z),\quad i \in [x,y]\,,
\end{equation}
with $\mathcal{C}^{(i)} \equiv \big(\ \alpha_{ix},\, 
\alpha_{iy},\, -\eta_{iy},\, \eta_{ix}\ \big)^\top$ 
the unknown coefficient vector. Note that the minus 
sign for $\eta_{xy}$ and $\eta_{yy}$ as well as the 
swapped order is simply a consequence of the differing 
convention for the $\eta$ tensor, which in our work 
is defined with respect to the current. The two 
portions, $\mathbf{E}^{(i)}$, of the left-hand vector 
are comprised of statistical moments of the EMF with 
the mean field and its gradients, that is,
\begin{equation}
  \mathbf{E}^{(i)} \equiv \Big(\ \big<\emf_i B_x\big>,\ \big<\emf_i B_y\big>,\ \big<\emf_i B'_x\big>,\ \big<\emf_i B'_y\big>\ \Big)^\top\,,
\end{equation}
where $B'_x\equiv \partial_z B_x$ and $B'_y\equiv 
\partial_z B_y$. Finally, the matrix $\mathcal{
\underline{M}}$, which is the same for the two sub 
portions, is given by
\begin{equation}
\mathcal{\underline{M}} \equiv  \left( \begin{array}{cccc}
\big<B_xB_x\big> & \big<B_xB_y\big> & \big<B_xB'_x\big> & \big<B_xB'_y\big>\\
\big<B_yB_x\big> & \big<B_yB_y\big> & \big<B_yB'_x\big> & \big<B_yB'_y\big>\\
\big<B'_xB_x\big> & \big<B'_xB_y\big> & \big<B'_xB'_x\big> & \big<B'_xB'_y\big>\\
\big<B'_yB_x\big> & \big<B'_yB_y\big> & \big<B'_yB'_x\big> & \big<B'_yB'_y\big>\\
\end{array}\right)\,.
\end{equation}
All entries are evaluated independently for each 
position $z$, as required for providing vertical 
profiles of the dynamo coefficients. To reduce the 
level of noise, we accumulate the data into coarser 
units in the vertical direction, typically by bunching 
together four or eight cells along $z$. In the above
expressions, angular brackets denote averaging over 
the data basis comprised by snapshots in time as well 
as $z$~positions that fall within the same vertical 
bin.

\begin{figure}
\centering
\includegraphics[width=\linewidth]{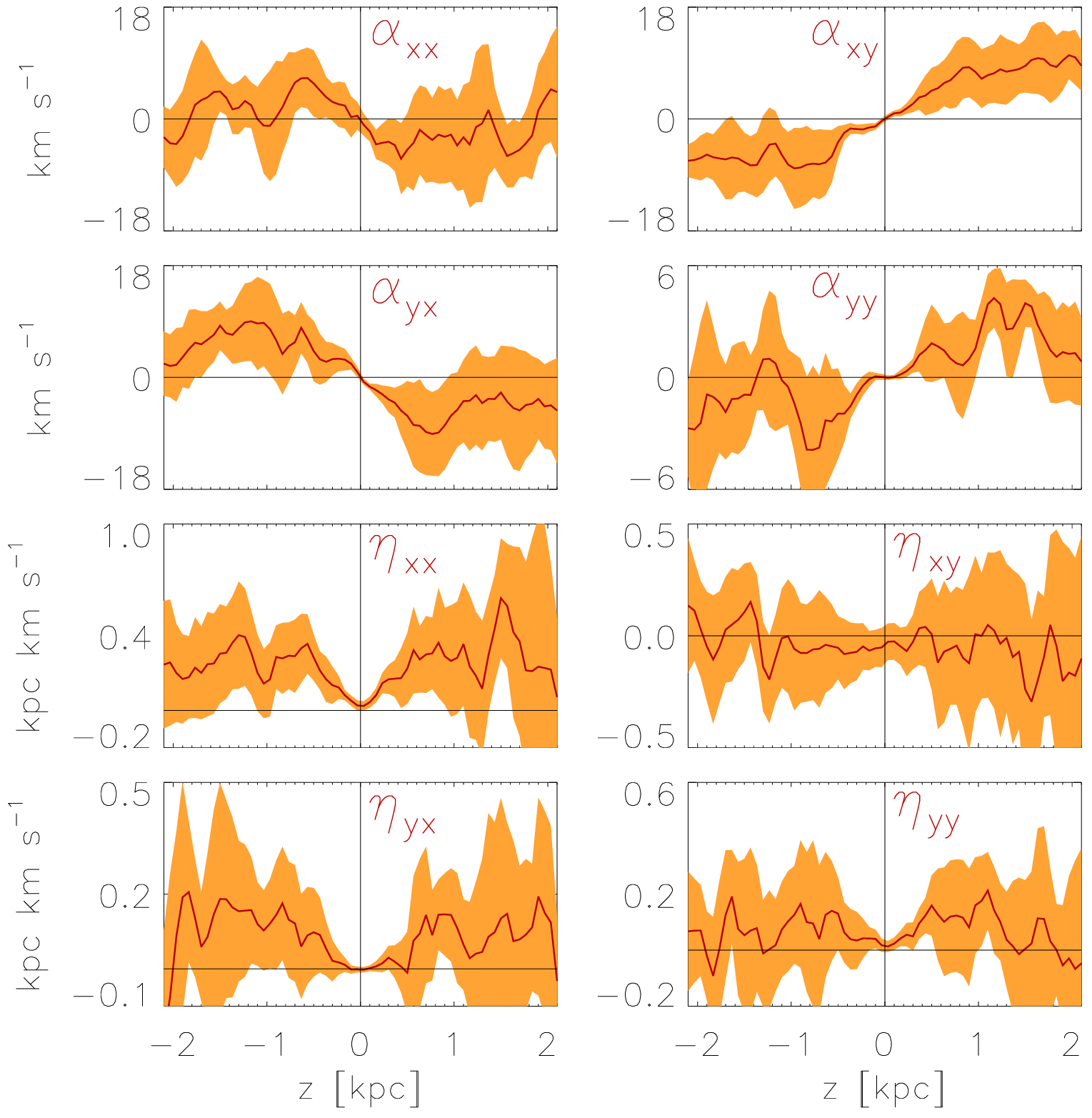}
\caption{Same as \fref{alpha_eta_slots} and 
	\fref{alpha_eta_tf}, but showing results for the 
	local regression method of {BS02}. Shaded 
	regions indicate the standard variations of the 
	coefficients when considering five separate 
	sub-intervals in time.}
\label{fig:alpha_eta_regression}
\end{figure}

As mentioned, we apply this method to the same raw data 
as is used for the SVD. The only difference is that the 
detrending in time was done by normalising with the 
instantaneous value of the horizontal field, $B_{\rm 
rms}(t) \equiv \sqrt{L_z^{-1}\,\int {\rm d}z \big(B_x^2(
z,t)+B_y^2(z,t)\big)}$ , rather than by subtracting an 
exponential growth factor, as was done for SVD. As 
mentioned before, the purpose of the detrending is to 
make sure that all data are contributing in a roughly 
similar manner, and small changes in the procedure were 
not found to have a significant impact.

Unsurprisingly, the results via the regression method, 
shown in \fref{fig:alpha_eta_regression}, agree markedly 
well with those derived via the SVD method (see 
\fref{alpha_eta_slots}). {The standard deviation is 
however somewhat larger than that obtained in the SVD 
analysis using 9 independent time series (see 
\fref{alpha_eta}).}


\bsp	
\label{lastpage}
\end{document}